\documentclass[a4paper,11pt]{article}
\pdfoutput=1 
\usepackage{jheppub} 
\usepackage{amsmath,amsfonts,amssymb,mathrsfs,graphicx,color,longtable,bm,wasysym}
\usepackage{hyperref}
\usepackage{graphicx}
\usepackage{array}
\usepackage{color}
\usepackage[usenames,dvipsnames,table]{xcolor}
\usepackage{tikz}
\usepackage[utf8]{inputenc}
\usepackage{epsfig,relsize,xspace}
\usepackage[T1]{fontenc}
\usepackage{bbold}
\usepackage{float}
\usepackage{amsmath}
\usepackage{empheq}
\usepackage{booktabs}
\usepackage{color}
\usepackage[utf8]{inputenc}
\usepackage{xspace}
\usepackage{scalerel}
\usepackage[most]{tcolorbox}
\usepackage{multirow}
\usepackage{scalefnt}
\usepackage{bold-extra}
\usepackage[shortlabels]{enumitem}
\usepackage[tikz]{bclogo}
\usepackage{subcaption}
\usepackage{tikz-feynman}
\usepackage{cancel}
\usepackage{verbatim}
\usetikzlibrary{arrows,shapes}
\usepackage{afterpage}
\usepackage{graphicx,grffile}
\usepackage{slashed}
\usepackage{soul}

\allowdisplaybreaks
 
\definecolor{nicered}{rgb}{0.7,0.1,0.1}
\definecolor{nicegreen}{rgb}{0.1,0.5,0.1}
\definecolor{niceblue}{rgb}{0.0,0.1,0.7}
\hypersetup{colorlinks,citecolor=niceblue,linkcolor=niceblue,urlcolor=niceblue}
                                      
\def \bm#1{\mbox{\boldmath$#1$\unboldmath}}
\def \beq{\begin{equation}}
\def \eeq{\end{equation}}
\def \bea{\begin{eqnarray}}
\def \eea{\end{eqnarray}}
\def \NLOplusPS {NLO$+$PS}
\def \NNLOplusPS {NNLO$+$PS}
\def \NNLOplusPSbm {NNLO$\bm{+}$PS}

\def \MiNNLOPS {MiNNLO$_{\rm PS}$}

\newcommand{\PYTHIA}[1]{{\tt Pythia~{#1}}\xspace}
\newcommand{\MiNLO}{MiNLO$^{\prime}$}

\begin{document}

\def\arraystretch{1.25}

\preprint{MPP-2022-53}

\title{NNLO event generation for $\bm{pp \to Zh \to \ell^+\ell^- b \bar b}$ production  in  the SM effective field theory}

\author[1]{Ulrich Haisch,}
\author[1]{Darren~J.~Scott,}
\author[1]{Marius Wiesemann,}
\author[1,2]{Giulia Zanderighi}
\author[1]{\\ and Silvia Zanoli}

\affiliation[1]{Max Planck Institute for Physics, \\ F{\"o}hringer Ring 6,  80805 M{\"u}nchen, Germany}
\affiliation[2]{Physik-Department, Technische Universit\"at M\"unchen, \\ James-Franck-Strasse 1, 85748 Garching, Germany}

\emailAdd{haisch@mpp.mpg.de}
\emailAdd{dscott@mpp.mpg.de}
\emailAdd{marius.wiesemann@mpp.mpg.de}
\emailAdd{zanderi@mpp.mpg.de}
\emailAdd{zanoli@mpp.mpg.de}

\abstract{We consider associated $Zh$ production with $Z \to \ell^+ \ell^-$ and $h \to b \bar b$ decays in hadronic collisions. In the framework of the Standard Model effective field theory~(SMEFT) we calculate the QCD corrections to
this process and achieve next-to-next-to-leading order plus parton shower~(\NNLOplusPS) accuracy using the \MiNNLOPS~method. This precision is obtained for a subset of six SMEFT operators, including  the corrections from effective Yukawa- and chromomagnetic dipole-type interactions. Missing higher-order QCD effects associated with the considered dimension-six operators are  estimated  to have a relative numerical impact of less than a percent on the total rate once existing experimental limits on the relevant Wilson coefficients are taken into account.  We provide a dedicated Monte~Carlo~(MC) code that evaluates the NNLO SMEFT corrections on-the-fly in the event generation. This MC~generator is used to study the numerical impact of \NNLOplusPS~corrections on the kinematic distributions in $pp \to Zh \to \ell^+ \ell^- b \bar b$  production employing simple SMEFT benchmark scenarios. We identify the invariant mass~$m_{b \bar b}$ of the two $b$-tagged jets as well as the three-invariant jet mass $m_{b \bar b  j}$ as  particularly interesting observables to study SMEFT effects. These~distributions receive contributions that change both their normalisation and shape with the latter modifications depending on the exact jet definition.  To our knowledge SMEFT effects of this type  have so far not  been discussed in the literature. The presented MC~generator can also serve as a starting point to obtain~\NNLOplusPS~accuracy for a suitable enlarged set of effective operators in the future.}

\maketitle

\section{Introduction}
\label{sec:introduction}

The Standard Model (SM) predicts that a Higgs boson with a mass of $125 \, {\rm GeV}$ decays with a branching ratio close to 60\% to a pair of bottom quarks.  The most sensitive production channels for detecting $h \to b \bar b$ decays at the LHC are despite their notably smaller cross sections with respect to gluon-gluon fusion Higgs production, the associated production of a Higgs boson with a $W$ or $Z$ boson ($Vh$), where the leptonic decay of the vector boson enables a clean selection. The $h \to b \bar b$ decay mode has been observed by both the ATLAS and the CMS collaboration in LHC~Run~II~\cite{ATLAS:2018kot,CMS:2018nsn}, and these measurements constrain the $h \to b \bar b$ signal strength in $V h$ production $\big ( \mu_{b \bar b}^{Vh} \big )$ to be SM-like within about~25\%, at the level of one standard deviation. With ATLAS and CMS being able to collect new data very soon in LHC~Run~III and with the high-luminosity upgrade~(HL-LHC) on the horizon, the precision of the $\mu_{b \bar b}^{Vh}$  measurements will  improve significantly with an ultimate projected HL-LHC accuracy  of $15 \%$ ($5\%$) in the case of $Wh$ ($Zh)$ production~\cite{ATLAS:2018jlh,CMS:2018qgz}.

Future LHC measurements of in particular $pp \to Zh \to \ell^+ \ell^- b \bar b$ production will thus allow to place strong constraints  on non-standard Higgs interactions. Assuming that any potential beyond-the-SM~(BSM) contributions arises only from particles with masses much heavier than the electroweak~(EW) scale, the  SM effective field theory~(SMEFT)~\cite{Buchmuller:1985jz,Grzadkowski:2010es,Brivio:2017vri} offers a largely model-independent, systematically improvable quantum-field theoretical framework for analyses of BSM effects in collider processes that can be used to probe Higgs production and decay. In fact, radiative corrections in the SMEFT to both $Vh$ production~\cite{Mimasu:2015nqa,Degrande:2016dqg,Alioli:2018ljm} and the $h \to f \bar f$ decays~\cite{Gauld:2015lmb,Gauld:2016kuu,Cullen:2019nnr,Cullen:2020zof} have been calculated. The recent paper~\cite{Bizon:2021rww} has furthermore studied QCD corrections to $pp \to Zh \to \ell^+ \ell^- b \bar b$ in the anomalous-coupling framework~(see also~\cite{Maltoni:2013sma,Greljo:2017spw} for earlier works on $pp \to Vh$ production in this context). The~existing studies for $Vh$ production have focused on the subset of higher-dimensional interactions that modify the couplings of the Higgs to EW gauge bosons achieving next-to-leading order~(NLO)~\cite{Mimasu:2015nqa,Degrande:2016dqg,Alioli:2018ljm,Maltoni:2013sma,Greljo:2017spw} and next-to-next-to-leading~order~(NNLO)~\cite{Bizon:2021rww}  in QCD perturbation theory, respectively, while in the case of $h \to b \bar b$ both NLO~QCD and NLO~EW corrections to the total  decay width have been calculated for the full set  of relevant dimension-six SMEFT~operators~\cite{Gauld:2015lmb,Gauld:2016kuu,Cullen:2019nnr}.  

The ultimate goal of  a  SMEFT calculation of the $pp \to Zh \to \ell^+ \ell^- b \bar b$  process consists in taking into account all relevant contributions of dimension-six operators to both $pp \to Zh$ and $h \to b \bar b$, and to combine the production and the decay processes into a Monte~Carlo~(MC) event generator that  consistently includes higher-order QCD and EW corrections at fixed order as well as parton-shower~(PS) effects. Within the~SM such computations have reached \NNLOplusPS~precision~\cite{Astill:2018ivh,Alioli:2019qzz,Bizon:2019tfo,Zanoli:2021iyp}, which means that they include NNLO~QCD corrections to $pp \to Zh \to \ell^+ \ell^- b \bar b$  production as well as the matching to PS~MC generators.  A dedicated PS~MC code  including both NLO QCD and EW corrections within the SM also exists~\cite{Granata:2017iod}. 

The intent of this work is to achieve \NNLOplusPS~accuracy in the SMEFT including all dimension-six operator insertions that contribute to the subprocesses $pp \to Zh$ and $h \to b \bar b$ directly in QCD. This requires to calculate QCD corrections for a subset of six SMEFT operators, including  effects associated to  effective Yukawa- and chromomagnetic dipole-type interactions. The inclusion of these contributions in the \NNLOplusPS~event generation yields an accurate description of the SMEFT effects in differential predictions for the full $pp \to Zh \to \ell^+ \ell^- b \bar b$ reaction.   Since the~distribution of the invariant mass~$m_{b \bar b}$ of the two $b$-tagged jets~($b$-jets)  receives contributions that change both its normalisation and shape  it turns out to provide a particularly  useful observable in the context of the~SMEFT.  The latter modifications depend on the~$b$-jet definition,  a feature that has not  been discussed in the SMEFT literature as far as we know.  Similar observations can be made for the three-jet invariant mass $m_{b \bar b j}$. As a by-product of our computations we are also able to extend the calculation of the  inclusive $h \to b \bar b$ decay rate in the SMEFT~\cite{Gauld:2016kuu} to the next order in~QCD for the case that bottom quarks are treated as massless. On the other hand, non-trivial EW corrections arising in the SMEFT, such as a modified~$hZZ$ coupling, are not taken into account in our analysis. Since~QCD and~EW SMEFT corrections to  $pp \to Zh \to \ell^+ \ell^- b \bar b$  approximately factorise, in the sense that the EW operators contributing to production play only a subleading role in the decay and vice versa, obtaining  \NNLOplusPS~accuracy for a suitable enlarged set of effective operators  should be rather straightforward. While~beyond the scope of this publication, we plan to return to  this problem in the future.

This manuscript is structured as follows: in Section~\ref{sec:preliminaries} we specify the subset of dimension-six operators of the full SMEFT Lagrangian that are relevant in the context of this article,  including a discussion of the normalisation chosen for the individual  effective interactions. Section~\ref{sec:calculation}~contains a brief description of the basic ingredients of the SMEFT calculations for $pp \to Zh$ and $h \to b \bar b$ and their combination and implementation in our \NNLOplusPS~event generator. The impact of the SMEFT corrections on kinematic distributions in $pp \to Zh \to \ell^+ \ell^- b \bar b$  production at  \NNLOplusPS~is presented in Section~\ref{sec:numerics} by using simple benchmark scenarios for the  Wilson coefficients. We conclude and present an outlook in~Section~\ref{sec:conclusions}. The~lenghty analytic expressions for the squared matrix elements that are relevant for our work are relegated to~Appendix~\ref{app:MEs},  while Appendix~\ref{app:others} contains numerical estimates of  higher-order QCD corrections associated to the subset of  the SMEFT operators that are considered in this paper.  The~discussed corrections have been neglected in our  phenomenological study because they all turn out to contribute less than a percent once existing experimental limits on the relevant Wilson coefficients are taken into account. 

\section{Preliminaries}
\label{sec:preliminaries}

In this article we consider the following set of dimension-six operators 
\begin{align} \label{eq:operators}
Q_{H \Box} &= ( H^\dagger H) \hspace{0.5mm} \Box \hspace{0.5mm} ( H^\dagger H ) \,, & 
Q_{HD} & = (H^\dagger D_\mu H )^\ast \hspace{0.5mm}  (H^\dagger D^\mu H )  \,, \nonumber \\[2mm]
Q_{bH} &= y_b \hspace{0.25mm}( H^\dagger H) \hspace{0.5mm} \bar q_L  \hspace{0.25mm}  b_R   \hspace{0.25mm} H  \,, &  
Q_{bG} & =  \frac{g_s^3}{(4 \pi)^2}  \hspace{0.5mm} y_b \hspace{0.5mm} \bar q_L  \sigma_{\mu \nu}  T^a \hspace{0.25mm}  b_R \hspace{0.25mm}  H \hspace{0.25mm}  G^{a,  \hspace{0.25mm} \mu \nu} \,, \\[2mm]
Q_{HG} &= \frac{g_s^2}{(4 \pi)^2} \hspace{0.5mm}  ( H^\dagger H) \hspace{0.5mm} G_{\mu \nu}^a \hspace{0.25mm}  G^{a,  \hspace{0.25mm} \mu \nu} \,, &  
Q_{3G} & = \frac{g_s^3}{(4 \pi)^2}  \hspace{0.5mm}  f^{abc} \hspace{0.25mm} G_{\mu}^{a,  \hspace{0.25mm} \nu} G_{\nu}^{b,  \hspace{0.25mm} \sigma}  G_{\sigma}^{c,  \hspace{0.25mm} \mu} \,, \nonumber 
\end{align} 
which appear in the full SMEFT Lagrangian 
\beq \label{eq:SMEFT}
{\cal L}_{\rm SMEFT} \supset \sum_{i} \frac{C_i}{\Lambda^2} \, Q_i \,.
\eeq
Here $\Box = \partial_\mu  \hspace{0.25mm}  \partial^\mu$, $\sigma_{\mu \nu} = i/2 \hspace{0.5mm} ( \gamma_\mu \gamma_\nu - \gamma_\nu \gamma_\mu)$ with $\gamma_\mu$ the usual Dirac matrices, $H$ denotes the SM Higgs doublet, $q_L$ is the left-handed third-generation quark doublet, $b_R$ is the right-handed bottom-quark singlet, while $g_s = \sqrt{4 \pi\alpha_s}$ and $G_{\mu \nu}^a$ denote the coupling constant and the field strength tensor of QCD, respectively. The definition of the covariant derivative  is  $D_\mu = \partial_\mu - i  \hspace{0.25mm}   g_s  \hspace{0.25mm}   G_\mu^a  \hspace{0.25mm}   T^a$  with $T^a$ being the $SU(3)$ generators and $f^{abc}$ denote the fully antisymmetric QCD structure constants. The bottom-quark Yukawa coupling is  defined as $y_b = \sqrt{2} \hspace{0.25mm} \bar m_b/v$, with the $\overline{\rm MS}$~bottom-quark mass $\bar m_b$ and the Higgs vacuum expectation value~(VEV) $v$, while $\Lambda$~denotes the new-physics mass  scale that suppresses the dimension-six operators~$Q_i$ entering~(\ref{eq:SMEFT}) and  $C_i$ are the corresponding Wilson coefficients. Notice finally that in the case of $Q_{bH}$ and $Q_{bG}$ the sum over  the hermitian conjugate in~(\ref{eq:operators}) is understood. 

The normalisations of the dimension-six operators introduced in~(\ref{eq:operators}) deserve some additional comments. First, the two mixed-chirality operators  $Q_{bH}$ and $Q_{bG}$ include a factor of $y_b$ which serves as an order parameter and explicitly appears in a broad class of ultraviolet~(UV) completions that match onto the set of operators in~(\ref{eq:operators}). See for example the discussions in~\cite{Giudice:2007fh,Elias-Miro:2013mua}. Second, the factors of $g_s$ and $1/(4 \pi)^2$ that arise in the definition of~$Q_{bG}$, $Q_{HG}$ and $Q_{3G}$ guarantee that the associated Wilson coefficients $C_{bG}$, $C_{HG}$ and~$C_{3G}$ are expected to  be   of~${\cal O} (1)$  in all weakly-coupled UV-complete extensions of the SM with new degrees of freedom and masses in the ballpark of $\Lambda$.\footnote{In the case of the operator $Q_{bG}$ the corresponding Wilson coefficient can also be of~${\cal O} (g^2/g_s^2)$ with $g$~a weak coupling   which implies that $C_{bG}$ is parametrically smaller than~${\cal O} (1)$. This happens when $Q_{bG}$ is generated by a EW and not a strong loop in the UV theory.} Notice that in the operator basis that has been employed in the NLO QCD calculation of the  inclusive $h \to b \bar b$ decay rate in the SMEFT~\cite{Gauld:2016kuu} a different normalisation is chosen for the operators $Q_{bH}$, $Q_{bG}$ and~$Q_{HG}$.  As a result, in this normalisation the corresponding Wilson coefficients are expected to be of size $C_{bH} = {\cal O} (y_b) = {\cal O} \big (10^{-2} \big )$, $C_{bG} = {\cal O} \big ( y_b \hspace{0.25mm} \alpha_s  /(4\pi) \big ) = {\cal O} \big ( 10^{-4} \big )$ and  $C_{HG} = {\cal O} \big ( \alpha_s  /(4\pi) \big ) = {\cal O} \big (10^{-2} \big )$ in weakly-coupled BSM theories and not $C_i = {\cal O} (1)$ as in the operator basis~(\ref{eq:operators}).  Our normalisation therefore has the merit that the suppression factors $y_b$ and $\alpha_s/(4 \pi)$ appear explicitly as order parameters which allows for a more explicit power counting in our SMEFT calculation of  QCD corrections to the  $pp \to Zh \to\ell^+ \ell^- b \bar b$ process.

\section{Calculation in a nutshell}
\label{sec:calculation}

In this section, we describe the different ingredients of the calculation of the QCD corrections to  the fully differential decay rate of $h \to b \bar b$ and the $Zh$ production cross section in the~SMEFT. Throughout this work, we use the five-flavour scheme and thus treat the bottom quark as massless both in the matrix elements and the phase-space integrals. The~bottom-quark Yukawa coupling is however taken to be non-zero. The explicit expressions for the non-trivial $h \to b \bar b$ squared matrix elements can be found in Appendix~\ref{app:MEs}. Moreover, we assume minimal-flavour violation~\cite{DAmbrosio:2002vsn} and set the Cabibbo-Kobayashi-Maskawa matrix element $V_{tb}$ to~unity. After~having discussed the anatomy of the SMEFT corrections to both  $pp\to Zh$ production and the $h \to b \bar b$ decay, we briefly mention the employed \NNLOplusPS~methods and explain how we apply them to~the~event generation of the $pp \to Zh \to\ell^+ \ell^- b \bar b$ process including SMEFT effects.
 
\subsection[{Factorisable contributions to the $h \to b \bar b$ decay}]{Factorisable contributions to the $\bm{h \to b \bar b}$ decay} 
\label{sec:fac}

Since the operators $Q_{H \Box}$, $Q_{HD}$ and  $Q_{bH}$ do not contain a gluon the associated SMEFT contributions  factorise to all orders in $\alpha_s$. As a result, the matrix elements proportional to the Wilson coefficients $C_{H \Box}$, $C_{HD}$ and  $C_{bH}$ can be obtained from the massless  NNLO calculation of the fully differential  $h \to b \bar b$ decay rate within the SM~\cite{Anastasiou:2011qx,DelDuca:2015zqa,Caola:2019pfz}  by the following simple replacement:
\beq \label{eq:factorising}
y_b^2 \to y_b^2 \hspace{0.75mm} \big ( 1 +  2   \hspace{0.125mm}  c_{\rm fac} \big )  \,,
\eeq
with 
\beq \label{eq:cfac}
c_{\rm fac} = c_{\rm kin} - c_{bH} \,, \qquad 
c_{\rm kin}   = \frac{v^2}{\Lambda^2}  \left [ C_{H \Box} - \frac{C_{HD}}{4}  \right ] \,, \qquad
c_{bH}   = \frac{v^2}{\Lambda^2}  \hspace{0.5mm}  {\rm Re} \left ( C_{bH} \right ) \,.
\eeq
Notice that the term $c_{\rm kin}$  arises  from  the canonical normalisation of the Higgs kinetic term in the presence of~$Q_{H \Box}$ and~$Q_{HD}$. The squared Higgs VEV entering~(\ref{eq:cfac}) is related  to the Fermi constant $G_F $ extracted from muon decay via $v^2 = 1/\big (\sqrt{2} \hspace{0.25mm} G_F \big)$.  EW tree-level corrections appearing in the SMEFT that modify the relation between the VEV and the Fermi constant $\big($see for example~\cite{Brivio:2017vri,Gauld:2015lmb}$\big)$ are very small  and hence neglected in our analysis. In practice, our implementation relies on the expressions for the squared matrix elements provided in~\cite{DelDuca:2015zqa} which also have been used in the publications~\cite{Astill:2018ivh,Bizon:2019tfo,Zanoli:2021iyp} to obtain \NNLOplusPS~predictions for $pp \to Zh \to \ell^+ \ell^- b \bar b$ production within the SM.

Given the simplicity of the replacement~(\ref{eq:factorising}) it is straightforward to obtain an analytic result for the factorisable corrections to the inclusive $h \to b \bar b$ decay rate due to $Q_{H \Box}$, $Q_{HD}$ and  $Q_{bH}$ up to NNLO in QCD. 
The leading-order~(LO) expression for the partial Higgs boson decay width to massless bottom-quark pairs within the SM is given by
\beq \label{eq:GammaSMLO}
\Gamma ( h \to b \bar b )_{\rm SM}^{\rm LO} = \frac{3 \hspace{0.25mm} y_b^2  \hspace{0.25mm} m_h}{16 \pi}  \,. 
\eeq
Using this expression the corresponding NNLO result takes the form~\cite{Gorishnii:1983cu,Gorishnii:1990zu,Gorishnii:1991zr}
\beq \label{eq:GammaSMNLO}
\Gamma ( h \to b \bar b )_{{\rm SM}}^{\rm NNLO} = \big ( 1 + \Delta \big)  \hspace{0.5mm} \Gamma ( h \to b \bar b )_{\rm SM}^{\rm LO}   \,, \qquad \Delta= \frac{\alpha_s}{\pi} \hspace{0.5mm} 5.67 + \left ( \frac{\alpha_s}{\pi} \right )^2  \hspace{0.5mm} 29.15 \,.
\eeq
Here both the bottom-quark Yukawa coupling $y_b$ and the strong coupling constant $\alpha_s$ are understood to be renormalised at the scale $m_h$ and we have assumed five active quark flavours to obtain the numerical results for the expansion coefficients appearing in $\Delta$. By~means of the replacement~(\ref{eq:factorising}) one then finds 
\beq \label{eq:GammaSMEFTNNLOfac}
\Gamma ( h \to b \bar b )_{{\rm SMEFT}}^{{\rm NNLO}, {\rm fac}} = \big ( 1 +  2   \hspace{0.125mm}  c_{\rm fac} \big )  \hspace{0.5mm}  \Gamma ( h \to b \bar b )_{{\rm SM}}^{\rm NNLO} \hspace{0.25mm}  \,,
\eeq
where the expressions for $c_{\rm fac}$ and $\Gamma ( h \to b \bar b )_{{\rm SM}}^{\rm NNLO}$ have already been given in~(\ref{eq:cfac}) and~(\ref{eq:GammaSMNLO}), respectively. Since in the limit of massless bottom quarks  the inclusive $h \to b \bar b$ decay rate is known to~${\cal O} (\alpha_s^4)$~\cite{Chetyrkin:1996sr,Baikov:2005rw,Herzog:2017dtz} an extension of~(\ref{eq:GammaSMEFTNNLOfac}) to~${\rm N}^4{\rm LO}$ in QCD would be possible. Such~higher-order QCD corrections are included in {\tt eHDECAY}~\cite{Contino:2013kra,Contino:2014aaa}, which allows to calculate the factorisable SMEFT corrections to $h \to b \bar b$ in the so-called strongly-interacting light Higgs or SILH~\cite{Giudice:2007fh} basis of dimension-six operators.  Since the aim of this paper is to consistently achieve \NNLOplusPS~accuracy for the  fully differential decay rate of $h \to b \bar b$ in the SMEFT, the result in~(\ref{eq:GammaSMEFTNNLOfac}) will however be sufficient for our purposes. 

\subsection[{Non-factorisable contributions to the $h \to b \bar b$ decay}]{Non-factorisable contributions to the $\bm{h \to b \bar b}$  decay} 
\label{sec:non}

\begin{figure}[t!]
\begin{center}
\includegraphics[width=0.75\textwidth]{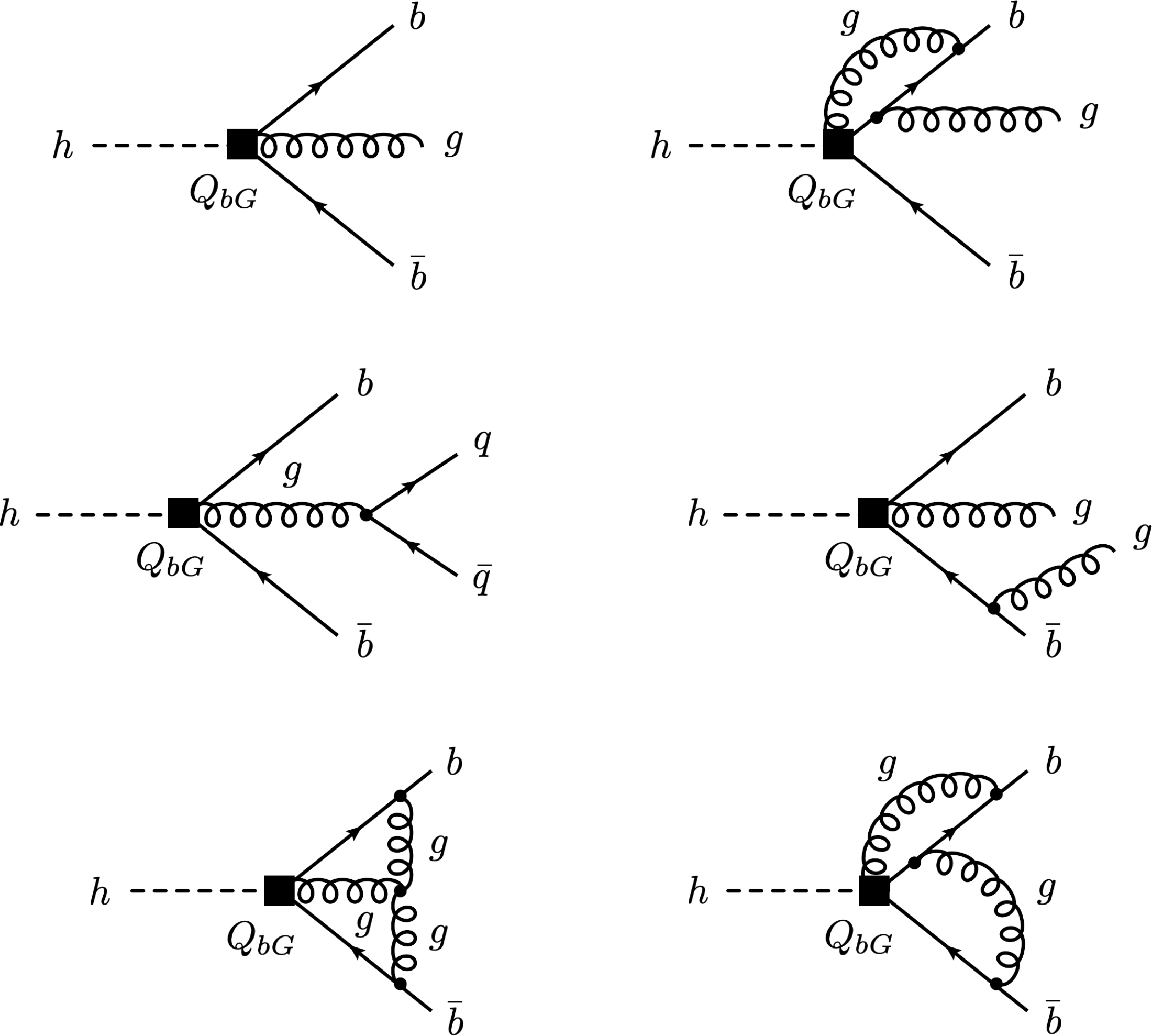}
\end{center}
\vspace{2mm} 
\caption{\label{fig:diagramsQbG} Examples of SMEFT contributions to the fully differential $h \to b \bar b$ decay rate involving an insertion of the  operator~$Q_{bG}$~(black square). The upper left~(right) diagram represents a tree-level (one-loop) contribution to the $h \to b \bar b g$ decay,  the center left~(right) diagram yields a tree-level contribution to the $h \to b \bar b q \bar q$ $\big($$h \to b \bar b gg$$\big)$ process, while  the lower diagrams contribute to the $h \to b \bar b$ amplitude at the two-loop level. Notice that the quark flavours~in the center left  $h \to b \bar b q \bar q$ diagram can be $q = u,d, s,c,b$ and that effective five-point $hb\bar b gg$ vertices  also contribute in the case of the $h \to b \bar b gg$ transition. See text for further~details. }
\end{figure}

Insertions of the operator $Q_{bG}$ lead to non-factorisable QCD corrections to the fully differential  $h \to b \bar b$ decay rate. Examples of the respective Feynman diagrams are shown in~Figure~\ref{fig:diagramsQbG}. The relevant squared matrix elements are found by interfering the SMEFT contributions with the corresponding SM amplitudes. The explicit expressions for the resulting squared matrix elements can be found in Appendix~\ref{app:MEs}. The leading contribution  arises at~${\cal O} \big ( y_b^2 \hspace{0.25mm} \alpha_s^2 \hspace{0.25mm}  C_{bG} \big )$  from the interference of the $h \to b \bar b g$ amplitude in the SMEFT~(upper left diagram) and the corresponding SM graphs. Notice that this real contribution is IR~finite and that the corresponding one-loop contribution to $h \to b \bar b$ is identical to zero since it only involves scaleless integrals. In the operator basis~(\ref{eq:operators}) the first non-factorisable QCD corrections involving the operator $Q_{bG}$ therefore appear at NNLO in QCD. In terms of the inclusive~LO~SM~decay rate~(\ref{eq:GammaSMLO}), we find by integrating~(\ref{eq:MEybdgLO}) over the  three-particle phase space the following compact expression:
\beq \label{eq:GammaSMEFTNNLOnonfac}
\Gamma ( h \to b \bar b )_{\rm SMEFT}^{{\rm NNLO}, {\rm non}} = \Delta_{\rm non} \hspace{0.5mm}  c_{bG} \hspace{0.5mm}  \Gamma ( h \to b \bar b )_{\rm SM}^{\rm LO} \,, \qquad
\Delta_{\rm non}  = \left ( \frac{\alpha_s}{\pi} \right )^2  \hspace{0.5mm}  \frac{m_h^2}{3 \hspace{0.25mm} v^2}  \,,  
\eeq
with 
\beq \label{eq:cbGdef}
c_{bG} = \frac{v^2}{\Lambda^2} \hspace{0.5mm} {\rm Re} \left ( C_{bG} \right ) \,. 
\eeq
We emphasise that when taking the difference in the normalisation of the operator~$Q_{bG}$ into account, the result~(\ref{eq:GammaSMEFTNNLOnonfac}) agrees with the expression derived in~\cite{Gauld:2016kuu} after taking  the limit~$m_b \to 0$. 

In Section~\ref{sec:preliminaries} we have argued that the Wilson coefficient of $Q_{bG}$ is expected to be of~${\cal O} (1)$ in a wide class of UV-complete theories if the operator is normalised as in~(\ref{eq:operators}).  While~there are strong bounds on the imaginary part of $C_{bG}$ from~the electric dipole moment of the neutron~\cite{Haisch:2021hcg}, the real part entering~(\ref{eq:cbGdef}) is at present only very weakly constrained by experiments~\cite{Hayreter:2013kba,Bramante:2014hua,Haisch:2021hcg}. In fact, values of $c_{bG} = {\cal O} (100)$ are compatible with all existing low- and high-energy data (see~Section~\ref{sec:numerics} for details), and therefore from a purely phenomenological point of view it is possible that   $\alpha_s/\pi \hspace{0.5mm} c_{bG} = {\cal O} (1)$  or even larger. In such a case the NNLO correction in~(\ref{eq:GammaSMEFTNNLOnonfac}) is numerically of the size of a NLO correction. To deal with this possibility we decided to include in our work also all contributions of~${\cal O} \big ( y_b^2 \hspace{0.25mm} \alpha_s^3 \hspace{0.25mm}  C_{bG} \big )$. As shown in~Figure~\ref{fig:diagramsQbG} there are three types of such contributions. There are virtual~(upper~right~diagram) or real corrections  to the Born-level process $h \to b \bar b g$, the latter of which lead to either $h \to b \bar b q \bar q$~(center~left~diagram)  or $h \to b \bar b g g$~(center~right~diagram).  Notice~that the  four-quark final state involves the quark flavours~$q = u,d, s,c,b$ and  that final-state configurations with two bottom quarks and two gluons receive contributions  from both effective $hb\bar b g$  and $hb\bar b gg$ vertices. Also the two-loop virtual corrections to the $h \to b \bar b$ decay~(lower~diagrams) contribute at~${\cal O} \big ( y_b^2 \hspace{0.25mm} \alpha_s^3 \hspace{0.25mm}  C_{bG} \big )$.  In~Section~\ref{sec:numerics} we will show that the inclusion of the N$^3$LO corrections associated to $Q_{bG}$ are phenomenologically relevant, which provides a further rational to incorporate  them in our \NNLOplusPS{} calculation of $pp \to Zh \to \ell^+ \ell^- b \bar b$ production.  

We now turn our attention to the matrix elements involving the insertion of an operator~$Q_{HG}$ or~$Q_{3G}$.  In the former case one finds that in the limit of a massless bottom quark the operator~$Q_{HG}$  does not contribute to the $h \to b \bar b$ decay. This is related to the fact that for $m_b = 0$ an insertion of~$Q_{HG}$  does not lead to amplitudes such as $h \to b \bar b$ where the final-state bottom quarks have mixed chiralities. This feature is well-known from the SM calculation of the inclusive $h \to b \bar b$ rate where amplitudes involving a $h \to gg$ subdiagram first contribute at NNLO in QCD~\cite{Chetyrkin:1995pd,Larin:1995sqc} and only if non-zero quark masses are considered. In~fact, the latter results receive Sudakov-like double logarithms of the form $\alpha_s^2 \ln^2 (m_b^2/m_h^2)$ from both real and virtual contributions. The same feature is present in the SMEFT calculations of $h \to f \bar f$ processes~\cite{Gauld:2016kuu,Cullen:2019nnr,Cullen:2020zof}. While bottom-quark mass effects in the fully differential SM $h \to b \bar b$ decay rate have been studied at NNLO~\cite{Bernreuther:2018ynm,Behring:2019oci,Behring:2020uzq,Somogyi:2020mmk}, it is not known  how to  resum the double-logarithmic  contributions in the case of the $h \to b \bar b$ decay.\footnote{In the case of the inclusive $h \to \gamma \gamma$ and $h \to gg$ decays it has been shown  how to resum  the leading double-logarithmic corrections~\cite{Akhoury:2001mz,Liu:2017vkm,Liu:2019oav,Wang:2019mym}.}  Likewise, it is also unknown how to correctly treat such terms in any of  the existing \NNLOplusPS~procedures. We therefore neglect all corrections associated to the insertion of~$Q_{HG}$, which is formally correct in the limit of massless bottom quarks. We show in Appendix~\ref{app:others} that the corrections to the inclusive  $\Gamma ( h \to b \bar b )$ decay rate proportional to $C_{HG}$ and at leading power in $m_b$ can be bounded in a model-independent fashion to the level of a few  permille.  Neglecting such effects in the calculation of the fully differential $\Gamma ( h \to b \bar b )$ decay rate can therefore also be assumed to be an excellent approximation. 

Employing the operator basis introduced in~(\ref{eq:operators}) insertions of $Q_{3G}$ induce  tree-level corrections to $h \to b\bar b gg$ and one-loop corrections to $h \to b \bar b g$.  After interfering these two types of contributions with the relevant SM amplitudes the resulting SMEFT corrections are proportional to~${\cal O} \big ( y_b^2 \hspace{0.25mm} \alpha_s^3 \hspace{0.25mm}  C_{3G} \big )$ and hence formally of N$^3$LO. Since  our goal is it to only attain \NNLOplusPS~precision we  neglect  corrections due to $Q_{3 G}$. In~Appendix~\ref{app:others} we present an estimate of  this type of N$^3$LO corrections that suggests that given the existing bounds on the Wilson coefficient~$C_{3G}$, SMEFT contributions of~${\cal O} \big ( y_b^2 \hspace{0.25mm} \alpha_s^3 \hspace{0.25mm}  C_{3G} \big )$ can indeed only have a very minor numerical impact on the $h \to b \bar b$ decay distributions.

\subsection[Contributions to  $pp \to Zh$]{Contributions to  $\bm{pp \to Zh}$ production}
\label{sec:Zh}

Let us first recall that the only effective dimension-six  EW interactions  that we are taking into account in our analysis are related to the  three operators $Q_{H \Box}$, $Q_{HD}$ and $Q_{bH}$ introduced in~(\ref{eq:operators}) and that otherwise we only consider effective interactions that directly induce QCD corrections. In~particular, dimension-six SMEFT operators that lead to non-trivial modifications of the $hZZ$ vertex are not considered in our work. 

With this simplification the dominant  corrections to $Zh$ production  in the SMEFT are associated to~$Q_{H \Box}$ and~$Q_{HD}$. These operators already provide a contribution at Born level which can  be obtained by the shift 
\beq \label{eq:ghZZ2shift}
g_{hZZ}^2 \to g_{hZZ}^2 \hspace{0.75mm} \big ( 1 +  2  \hspace{0.125mm} c_{\rm kin} \big )  \,,
\eeq
where $g_{hZZ} = 2 \hspace{0.125mm} m_Z^2/v$ denotes the SM coupling between a Higgs and two $Z$ bosons with $m_Z$ the $Z$-boson mass.  The coefficient $c_{\rm kin}$ has been defined in~(\ref{eq:cfac}). Applying the rescaling~(\ref{eq:ghZZ2shift}) to the inclusive $Zh$ production cross section in the SM one obtains the following formula  
\beq \label{eq:sigmaZhSMEFTNNLO}
\sigma ( p p \to Zh )_{\rm SMEFT}^{\rm NNLO}= \big ( 1 +  2  \hspace{0.125mm} c_{\rm kin} \big )  \hspace{0.5mm}  \sigma ( p p \to Zh )_{\rm SM}^{\rm NNLO} \,, 
\eeq
for the SMEFT corrections to the cross section up to NNLO. An analogous expression also holds at the differential level. 

Let us also discuss the role of operators other  than $Q_{H \Box}$ and~$Q_{HD}$ in associated Higgs production with a vector boson. For massless bottom quarks neither $Q_{bH}$ nor $Q_{bG}$ furnishes a non-zero contribution to $pp \to Zh$ to all orders in the strong coupling constant $\alpha_s$ due to the mixed-chirality nature of the two operators. Insertions of $Q_{HG}$ lead to  a tree-level and an one-loop correction to  $q \bar q \to Zhg$  and $q \bar q \to Zh$, respectively.  After interfering these amplitudes with their SM~counterparts one obtains a contribution that is of~${\cal O} (\alpha_s^2 \hspace{0.5mm} C_{HG})$ with respect to the LO~$Zh$ production cross section within the SM. While these corrections are therefore formally needed to achieve \NNLOplusPS~accuracy, we will show in Appendix~\ref{app:others} that the contributions proportional to $C_{HG}$ cannot exceed the level of a few permille. This renders them irrelevant for all practical purposes and we hence neglect them. We note that by using the results of~\cite{Brein:2011vx} their inclusion would be quite straightforward, but given  their subleading impact we refrain from doing so. Finally, the operator $Q_{3G}$ gives rise to tree-level (one-loop) corrections to  $q \bar q \to Zhgg$~($q \bar q \to Zhg$). Both contributions are of ${\cal O} (\alpha_s^3 \hspace{0.5mm} C_{3G})$ and therefore  we do not take them into account in our analysis. In~Appendix~\ref{app:others} we nevertheless estimate their potential size and show them to be negligible for our purposes. We finally add that  contributions to $gg \to Zh$ production, which is loop suppressed in the SM, do~either vanish in the limit $m_b = 0$ as for $Q_{bH}$ and~$Q_{bG}$  or are at least of ${\cal O} ( \alpha_s^3)$ like in the case of $Q_{HG}$ and~$Q_{3G}$. Contributions to the  $gg \to Zh$ process arising from~(\ref{eq:operators}) are therefore phenomenologically irrelevant.

\subsection[\NNLOplusPS~calculation and MC implementation]{\NNLOplusPSbm~calculation and MC implementation}

In the following, we discuss how the fixed-order QCD results  including the set of dimension-six operators in~(\ref{eq:operators}) are implemented  into the \NNLOplusPS~accurate $h \to b \bar b$ generator developed in~\cite{Bizon:2019tfo}. We then describe how the $h \to b \bar b$  decay events can be combined with  Higgs production events from any Higgs production mode, including SMEFT effects consistently.  While in this paper we consider the case of associated $Zh$ production, the modifications to the formulae required for other processes are straightforward as long as one works in the narrow width approximation for the Higgs propagator. 

We start our discussion by briefly recalling how \NNLOplusPS~accurate results for the  $h \to b \bar b$ decay can be achieved in the SM. The simplest way to obtain~\NNLOplusPS~precision is to compute  decay events by applying the so-called \MiNLO~method~\cite{Hamilton:2012np,Hamilton:2012rf}. The weights are then rescaled to the NNLO accurate inclusive $h\to b\bar b$ decay rate $\Gamma (h \to b \bar b)_{\rm SM}^{\rm NNLO}$ by multiplying the event weights with the ratio $\Gamma (h \to b \bar b)_{\rm SM}^{\rm NNLO}/\Gamma (h \to b \bar b)_{\rm SM}^{{\rm MiNLO}^\prime}$, where the denominator is directly computed from the \MiNLO~events by summing their weights. As a variant, it is also possible to restrict the NNLO correction to a certain phase space region. In particular, one can use a rescaling factor that tends to one in the region where the decay events involve hard radiation, while requiring the integral of the rescaled events  to reproduce exactly $\Gamma (h \to b \bar b)_{\rm SM}^{\rm NNLO}$. In~this work, we follow such  an approach and specifically employ the procedure described in Section~2.2 of the article~\cite{Bizon:2019tfo}. 

As discussed  in Sections~\ref{sec:fac} and \ref{sec:non}, the SMEFT corrections to $h \to b \bar b$ split into factorisable and non-factorisable corrections. The factorisable SMEFT effects can be taken into account by using an analogous reweighting to~(\ref{eq:GammaSMEFTNNLOfac}) and applying it to the \NNLOplusPS~accurate $h \to b \bar b$ events, i.e.~by~computing $\big(1+2\hspace{0.125mm} c_{\rm fac}\big) \hspace{0.25mm} \Gamma (h \to b \bar b)_{\rm SM}^{\rm NNLO}/\Gamma (h \to b \bar b)_{\rm SM}^{{\rm MiNLO}^\prime}$ when reweighting the \MiNLO~events. As far as non-factorisable corrections are concerned, they are separately IR~finite, as pointed out already before. In particular, the leading contribution including $Q_{bG}$ is of ${\cal O} (\alpha_s^2)$ and contributes to the squared matrix element of $h\to b\bar bg$, cf.~(\ref{eq:MEybdgLO}), which features no divergence when the gluon is unresolved and will be referred to as ${\rm R}$ hereafter. Hence, the integration over the three-body phase space is IR~finite without the need to apply the \MiNLO~method, at variance with the  \NNLOplusPS~calculation of the factorisable part. At~${\cal O} (\alpha_s^3)$ the one-loop virtual and real corrections to the $h \to b\bar bg$ process are included, just as in a standard  POWHEG~\NLOplusPS~calculation~\cite{Nason:2004rx,Frixione:2007vw}. In the following, we will refer to these contributions as~${\rm RV}$ and~${\rm RR}$, respectively. Since their sum is IR~finite, these corrections yield a contribution to the inclusive $h \to b \bar b$ decay rate when integrated over the three- and four-body phase space, respectively, and added together.  To deal with the soft and collinear singularities of the real contributions and to cancel the IR~poles of the one-loop virtual corrections, cf.~(\ref{eq:MEybdg10}), we exploit the general implementation of the Frixione-Kunszt-Signer~(FKS) subtraction~\cite{Frixione:1995ms,Frixione:1997np}  within the~{\tt POWHEG-BOX} framework~\cite{Alioli:2010xd}. To~this end, the full {\tt POWHEG-BOX} machinery is used that automatically builds the soft and collinear counterterms and remnants, and also checks the behaviour in the soft and collinear limits of the real squared matrix elements against their soft and collinear approximations. Finally, also the two-loop virtual corrections to $h \to b \bar b$ including $Q_{bG}$ contribute to the non-factorisable SMEFT effects at~${\cal O} (\alpha_s^3)$. These corrections, called~${\rm VV}$ in what follows, are IR finite by themselves, cf.~(\ref{eq:hbb2loop}), and after integrating them over the two-particle phase space we find  the compact expression 
\beq \label{eq:da2loop}
\Gamma ( h \to b \bar b )_{\rm SMEFT}^{ {\rm non},{\rm VV}} = \frac{\alpha_s}{\pi} \hspace{0.125mm} \frac{49}{12} \hspace{0.5mm} \Delta_{\rm non} \hspace{0.5mm}  c_{bG} \hspace{0.5mm}  \Gamma ( h \to b \bar b )_{\rm SM}^{\rm LO} \,,
\eeq
with $\Delta_{\rm non}$ and $c_{bG}$ defined as in~(\ref{eq:GammaSMEFTNNLOnonfac}) and (\ref{eq:cbGdef}), respectively. The  result in~(\ref{eq:da2loop}) can simply be added to the factorisable SMEFT contribution to  the $h \to b \bar b$ decay.

To explain how we combine production and decay events in the SMEFT to obtain $pp \to Zh \to \ell^+ \ell^- b \bar b$ events, it is useful to first recall how such a combination can be achieved within the SM. In this case, the Higgs boson in each $Zh$ production event  is replaced by the possible decay products  (i.e.~$b \bar b g$, $b \bar b q \bar q$ or   $b \bar b gg$) taken from a given decay event. Working in the narrow width approximation, the weight $w_{\rm full}^{\rm SM}$ of the  full SM event is then calculated as 
\beq  \label{eq:fullweight}
w_{\rm full}^{\rm SM} = \frac{w_{\rm prod}^{\rm SM}  \, w_{\rm dec}^{\rm SM}}{\Gamma_h^{\rm SM}} \,, 
\eeq
where $w_{\rm prod}^{\rm SM}$ denotes  the weight of the production event obtained at \NNLOplusPS~using the \MiNNLOPS~$pp \to Zh$  generator presented in~\cite{Zanoli:2021iyp}, while $w_{\rm dec}^{\rm SM}$ represents  the weight of the decay event computed at \NNLOplusPS~precision employing  the  \MiNLO~method and procedure detailed in the work~\cite{Bizon:2019tfo}. The~total decay width $\Gamma_h^{\rm SM}$ of the $125 \, {\rm GeV}$ SM Higgs boson   serves as an input. The~Les~Houches event~(LHE) file also stores, for each event, the value of the hardest radiation allowed by the shower (i.e.~the scale {\tt scalup}). We record in the combined LHE file the value of {\tt scalup$_{\tt prod}$} for the production process and, once the event is passed to the PS, we recompute on-the-fly the value of {\tt  scalup$_{\tt dec}$} for the specific decay kinematics. We then generate emissions of the Higgs decay in all the available phase space and after the shower is complete we check whether the hardness of the splittings is below the veto scale {\tt  scalup$_{\tt dec}$}. If this is not the case we attempt to shower the event again until the condition is met. Additional details of this procedure are given in Section~3.2 of the paper~\cite{Bizon:2019tfo}. 

\begin{figure}[t!]
\begin{center}
\includegraphics[width=0.75\textwidth]{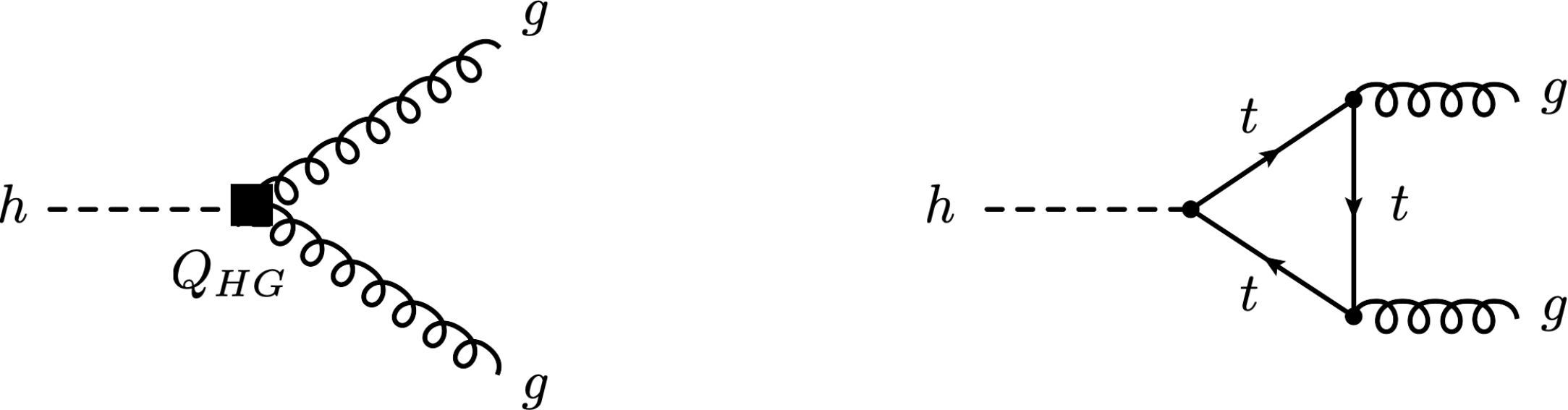}
\end{center}
\vspace{0mm} 
\caption{\label{fig:diagramswidth}  Illustration of the interference contribution that leads to the SMEFT correction proportional to $c_{HG}$ in the partial decay width $h \to g g$ as given in~(\ref{eq:GammahSMEFT}). The left and right diagram represent the tree-level SMEFT with an $Q_{HG}$ insertion~(black square) and the one-loop SM contribution involving a top-quark loop, respectively. Light-quark loops also appear in the~SM, but since  these corrections vanish identically for $m_q = 0$, they turn out to be numerically insignificant and hence are ignored in our study.  }
\end{figure}

From the above discussion one can deduce that \NNLOplusPS~accurate differential cross sections for the full process $pp \to Zh \to \ell^+ \ell^- b \bar b$ in the SMEFT can be obtained by combining production and decay as follows
\beq \label{eq:blackmagic}
\begin{split}
d\sigma_{\text{\NNLOplusPS}} & = \big ( 1 +  2  \hspace{0.125mm} c_{\rm kin} \big )^2  \, \Bigg \{  \left [ 1- 2 \hspace{0.125mm} c_{bH} + \frac{\Gamma ( h \to b \bar b )_{\rm SMEFT}^{{\rm non},{\rm VV}} }{\Gamma ( h \to b \bar b )_{{\rm SM}}^{\rm NNLO} } \right ] \hspace{0.25mm} d \sigma^{\rm SM}_{\text{\NNLOplusPS}} \\[2mm] 
& \hspace{2.75cm} \, + d\sigma^{\rm non, {\rm R}+{\rm RV}+{\rm RR}}_{\text{\NNLOplusPS}} \hspace{0.25mm} \Bigg \} \; \frac{\Gamma_h^{\rm SM}}{\Gamma_h^{\rm SMEFT}}\,,
\end{split}
\eeq 
where we have factorised the term $ \big ( 1 +  2  \hspace{0.125mm} c_{\rm kin} \big )^2$ that arises from the canonical normalisation of the Higgs kinetic term.   The results for $\Gamma ( h \to b \bar b )_{{\rm SM}}^{\rm NNLO}$ and $\Gamma ( h \to b \bar b )_{\rm SMEFT}^{{\rm non},{\rm VV}}$ can be found in~(\ref{eq:GammaSMNLO}) and (\ref{eq:da2loop}), $d \sigma^{\rm SM}_{\text{\NNLOplusPS}}$ represents the \NNLOplusPS~accurate differential cross section obtained interfacing the \MiNNLOPS~calculation of the $pp \to Zh$ generator~\cite{Zanoli:2021iyp} with the reweighted \MiNLO~calculation of the $h \to  b \bar b$ generator~\cite{Bizon:2019tfo}, while $d\sigma^{\rm non, {\rm R}+{\rm RV}+{\rm RR}}_{\text{\NNLOplusPS}}$ includes the non-factorisable corrections~${\rm R}$, ${\rm RV}$ and~${\rm RR}$ that we compute as discussed above. Notice~that both $\Gamma ( h \to b \bar b )_{\rm SMEFT}^{ {\rm non},{\rm VV}}$ and $d\sigma^{\rm non, {\rm R}+{\rm RV}+{\rm RR}}_{\text{\NNLOplusPS}}$  depend linearly on the Wilson coefficient~$c_{bG}$.  We stress that the advantage of expressing~(\ref{eq:blackmagic}) in this form  lies in the fact that arbitrary combinations of Wilson coefficients can be obtained without recalculating any of the individual cross sections. Thus, variations of the Wilson  coefficients can be obtained a posteriori.

The last factor in~(\ref{eq:blackmagic}) takes into account that the total decay width of the Higgs boson that appears  in~(\ref{eq:fullweight}) is modified by SMEFT effects. In our implementation we employ the following result
\begin{equation} \label{eq:GammahSMEFT}
\begin{split}
\Gamma_h^{\rm SMEFT} & =   \big ( 1 +  2  \hspace{0.125mm} c_{\rm kin} \big ) \Bigg [ \, \Gamma_h^{\rm SM} - \big ( 2  \hspace{0.25mm} \Delta  \hspace{0.25mm}  c_{bH} -    K_{bG} \hspace{0.5mm}  \Delta_{\rm non} \hspace{0.25mm} c_{bG}  \big )  \hspace{0.5mm}\Gamma ( h \to b \bar b )_{\rm SM}^{\rm LO} \\[2mm]
& \hspace{2.5cm} + 6 \hspace{0.25mm} K_{HG}  \hspace{0.25mm} c_{HG} \hspace{0.25mm}   \Gamma ( h \to g g )_{\rm SM}^{\rm LO} \,  \Bigg ] \,. 
\end{split}
\end{equation}
The LO expression for the partial decay width for $h \to b \bar b$ can be found in~(\ref{eq:GammaSMLO}) and the corrections proportional to $\Gamma ( h \to b \bar b )_{\rm SM}^{\rm LO}$  have  been included  in~(\ref{eq:GammahSMEFT}) in the approximation that treats the bottom quark as strictly massless.  The relevant correction factors $\Delta$ and $\Delta_{\rm non}$ have  been  defined in~(\ref{eq:GammaSMNLO}) and~(\ref{eq:GammaSMEFTNNLOnonfac}), respectively.  The~multiplicative factor $K_{bG}$ encodes the QCD corrections up to N$^3$LO related to the  $Q_{bG}$ contribution  to the partial decay width of $h \to  b \bar b$. The used numerical value of $K_{bG} = 1.622$  follows from the semi-analytic formula given below in~(\ref{eq:GammaSMEFTN3LO}). The term in $\Gamma_h^{\rm SMEFT}$ proportional to 
\beq \label{eq:cHGdef}
 c_{HG} = \frac{v^2}{\Lambda^2} \hspace{0.5mm} C_{HG} \,,
\eeq  
encodes the SMEFT corrections to the partial decay width of $h \to g g$ involving a single insertion of the operator~$Q_{HG}$. The relevant diagrams are shown in Figure~\ref{fig:diagramswidth}. These terms start at~${\cal O} (\alpha_s^2 \hspace{0.25mm} C_{HG})$ since 
\beq \label{eq:GammahggLOSM}
\Gamma ( h \to g g )^{\rm LO}_{\rm SM} = \frac{\alpha_s^2\hspace{0.25mm}  m_h^3}{72 \hspace{0.25mm} \pi^3 \hspace{0.25mm} v^2} \,.
\eeq
This result holds  in the limit of an infinite top-quark mass, which is a sufficiently accurate approximation for our purpose. To include higher-order effects proportional to $C_{HG}$ in an approximate way we have incorporated the factor $K_{HG} = 1.844$ in~(\ref{eq:GammahSMEFT}). This $K$-factor  includes QCD corrections up to~N$^4$LO~\cite{Herzog:2017dtz}. Notice finally that~(\ref{eq:GammahSMEFT}) contains only QCD corrections related to the dimension-six SMEFT operators introduced in~(\ref{eq:operators}).\footnote{By adding the term $\Delta \Gamma_h^{\rm SMEFT} = -  \big ( 1 +  2  \hspace{0.125mm} c_{\rm kin} \big )  \sum_{q=u,d,s,c} \big ( 2  \hspace{0.25mm} \Delta  \hspace{0.25mm}  c_{qH} -   K_{qG} \hspace{0.25mm}  \Delta_{\rm non} \hspace{0.25mm} c_{qG}  \big )  \hspace{0.5mm}\Gamma ( h \to q \bar q )_{\rm SM}^{\rm LO} $ to~(\ref{eq:GammahSMEFT}) with   $K_{qG} = 1.622$  it is straightforward to include the dominant SMEFT corrections to the  partial decay widths for~$h \to q \bar q$.}

We finally recall from the discussion in Section~\ref{sec:Zh} that if one considers associated $Zh$ production alone, the SMEFT production weight  would be modified by a universal factor leading to
\beq \label{eq:facme} 
w_{\rm prod}^{\rm SMEFT} = \left ( 1 +  2 \hspace{0.125mm} c_{\rm kin} \right ) w_{\rm prod}^{\rm SM}  \,.
\eeq
The multiplicative factor $ \left ( 1 +  2 \hspace{0.125mm} c_{\rm kin} \right )$, however,  cancels against the same overall factor appearing in the total decay width of the Higgs boson when production and decay events are combined. This cancellation becomes manifest in~(\ref{eq:blackmagic}) when~(\ref{eq:GammahSMEFT}) is used. If, on  the other hand, one considers  production processes other than $pp \to Zh$, where SMEFT effects give rise to non-universal corrections to the production process, some of the formulae given  in this subsection need to be modified accordingly.

\section{Phenomenological analysis}
\label{sec:numerics}

In the following  we present \NNLOplusPS~accurate results for $pp \to Zh \to \ell^+ \ell^- b \bar b$ production at the LHC with a centre-of-mass energy of $\sqrt{s} =13 \, {\rm TeV}$ including the SMEFT effects discussed in Section~\ref{sec:calculation}. All SM input parameters  are taken from the most recent PDG~review~\cite{Zyla:2020zbs}. In~particular, we use $G_F = 1.166379 \cdot 10^{-5} \, {\rm GeV}^{-2}$, $m_Z = 91.1876 \,  {\rm GeV}$, $\Gamma_Z^{\rm SM} = 2.4952 \, {\rm GeV}$, $m_h = 125.09 \, {\rm GeV}$ and $\Gamma_h^{\rm SM} = 4.1 \, {\rm MeV}$. The Higgs VEV is calculated using $v = 2^{-1/4} \hspace{0.25mm} G_F^{-1/2} = 246.22 \, {\rm GeV}$, while the value of the electromagnetic coupling and the weak mixing angle is derived as $\alpha =\sqrt{2} \hspace{0.25mm} G_F \hspace{0.25mm} m_W^2 \sin^2 \theta_w/\pi = 1/132.184$ and $\sin^2 {\theta_w} = 1-{m_W^2}/{m_Z^2} = 0.2230$ employing $m_W = 80.379 \, {\rm GeV}$.  In the case of the bottom-quark mass we use the $\overline {\rm MS}$ mass $\bar m_b (\bar m_b)  = 4.18 \, {\rm GeV}$ as input which leads to $\bar m_b (m_h) = 2.79 \, {\rm GeV}$ and  $y_b (m_h)= \sqrt{2} \hspace{0.25mm} \bar m_b (m_h)/v = 1.60 \cdot10^{-2}$. {\tt NNPDF31\_nnlo\_as\_0118} parton distribution functions~(PDFs)~\cite{Ball:2017nwa} with $\alpha_s (m_Z) = 0.1180$ giving rise to $\alpha_s (m_h) = 0.1127$ are employed in our MC~simulations and events are showered with~\PYTHIA{8}~\cite{Sjostrand:2014zea} utilising the Monash tune~\cite{Skands:2014pea}.  Effects~from hadronisation, underlying event modelling or  QED effects in the shower are not included. 

We study two SMEFT benchmark scenarios. In both cases we set the Wilson coefficients $c_{\rm kin}$, $c_{HG}$ and $c_{3G}$  to zero --- see~(\ref{eq:cfac}) and (\ref{eq:cHGdef}). In the first scenario, we choose $c_{bH} \neq 0$ and  $c_{bG} = 0$, while in the second one we take  $c_{bH} = 0$ and $c_{bG} \neq 0$. Before specifying our benchmark settings for the two parameters, let us summarise the existing experimental constraints on~$c_{bH}$ and $c_{bG}$.  In the former case, we take the result from a recent global SMEFT fit~\cite{Ellis:2020unq} that includes~34 dimension-six operators. In the normalisation of~(\ref{eq:operators}) and~(\ref{eq:cfac}), this work provides  the following marginalised 95\%~confidence level~(CL) bound
\beq \label{eq:cbH95CL}
c_{bH} \in [-0.13, 0.20] \,.
\eeq
In the case of $c_{bG}$ one instead has 
\beq \label{eq:cbG95CL}
c_{bG} \in [-438, 438] \,, 
\eeq
at 95\%~CL, which has been obtained in~\cite{Haisch:2021hcg} from an analysis of the transverse momentum~($p_T$) spectrum of $Z$-boson production in association with $b$-jets as measured by  ATLAS  in LHC~Run~II~\cite{ATLAS:2020juj}. We add that the nominal strongest bound on~$c_{bG}$ has been derived in~\cite{Haisch:2021hcg} and relies  on high-mass $b$-jet pair production ATLAS data~\cite{ATLAS:2019fgd}. While this search imposes $c_{bG} \in [-149, 149]$ it is not clear to which extent a SMEFT treatment is trustworthy in this high-mass region, so that we do not use it here. The present bounds from Higgs physics on the Wilson coefficient~$c_{bG}$~\cite{Hayreter:2013kba,Bramante:2014hua} are weaker than the constraint given in~(\ref{eq:cbG95CL}). 

\subsection[Inclusive $h \to b \bar b$ decay in the SMEFT]{Inclusive $\bm{h \to b \bar b}$ decay in the SMEFT}

As a first application of the calculations outlined in Section~\ref{sec:calculation}  we extend the results for the inclusive $h \to b \bar b$ decay width in the SMEFT presented in~\cite{Gauld:2016kuu}  to the  N$^3$LO level in~QCD for massless bottom quarks.  In terms of the LO SM inclusive decay width~(\ref{eq:GammaSMLO}) we~find 
\beq \label{eq:GammaSMEFTN3LO}
\begin{split}
\Gamma (h \to b \bar b)_{\rm SMEFT}^{\text{N$^3$LO}} & = \Bigg \{  \left (1 + 2 \hspace{0.125mm}  c_{\rm fac} \right ) \left [ 1 + \frac{\alpha_s}{\pi}  \hspace{0.5mm} 5.67 + \left ( \frac{\alpha_s}{\pi} \right )^2  \hspace{0.5mm} 29.15 + \left ( \frac{\alpha_s}{\pi} \right )^3 \hspace{0.25mm} 41.76 \right ] \\[2mm] 
& \phantom{xxx} + \left ( \frac{\alpha_s}{\pi} \right )^2  \frac{m_h^2}{3 \hspace{0.25mm} v^2}  \left [ 1 +  \frac{\alpha_s}{\pi}  \hspace{0.5mm} 17.32  \right ]  c_{bG} \Bigg \} \; \Gamma (h \to b \bar b)_{\rm SM}^{\text{LO}}  \,,
\end{split}
\eeq
if the renormalisation scale is identified with the Higgs boson mass by setting $\mu = m_h$. Notice that  the~${\cal O} (\alpha_s^3)$ corrections proportional to $\left (1 + 2 \hspace{0.125mm}  c_{\rm fac} \right )$ are known from the SM calculation of the $h \to b \bar b$ decay~\cite{Chetyrkin:1996sr,Baikov:2005rw,Herzog:2017dtz}.  The  result for the~${\cal O} (\alpha_s^3)$ correction proportional to~$c_{bG}$ is instead new and given here for the first time. Notice that the latter terms enhance the non-factorisable contribution due to $Q_{bG}$ by around~60\%, which provides a clear motivation to incorporate them in our \NNLOplusPS~generator. The corrections associated with $Q_{HG}$ and~$Q_{3G}$ are very small and not included in~(\ref{eq:GammaSMEFTN3LO}). See~the discussion in Appendix~\ref{app:others}. 

Using~(\ref{eq:GammaSMEFTN3LO}) one can now study the possible numerical impact of the set of dimension-six operators introduced in~(\ref{eq:operators}). Allowing for instance $c_{bH}$~($c_{bG}$) to vary within its experimentally allowed 95\%~CL range in eqs.~(\ref{eq:cbH95CL}) and (\ref{eq:cbG95CL}), while setting all other Wilson coefficients to zero, leads~to the following relative shifts in the inclusive $h \to b  \bar b$ decay width:
\beq \label{eq:max}
\frac{\Gamma (h \to b \bar b)_{\rm SMEFT}^{\text{N$^3$LO}} }{\Gamma (h \to b \bar b)_{\rm SM}^{\text{N$^3$LO}}}-1  \in  \left \{ \begin{matrix}  [-39,26] \%  & \text{for}~(\ref{eq:cbH95CL})~\text{and}~c_{\rm kin}= c_{bG}=0 \,, \\[2mm]  [-6.3,6.3] \%  &  \text{for}~(\ref{eq:cbG95CL})~\text{and}~c_{\rm kin}= c_{bH}=0 \,. \end{matrix} \right .
\eeq
Thus, there is a hierarchy between the possible SMEFT effects in the $h \to b \bar b$ decay rate,  with the non-factorisable contributions due to $Q_{bG}$ being smaller by a factor of~${\cal O} (5)$ than the factorisable corrections that are associated to $Q_{bH}$ (as well as  $Q_{H\Box}$ and $Q_{HD}$). The~SMEFT~QCD corrections in~(\ref{eq:GammaHG95CL}) and~(\ref{eq:Gamma3G95CL}) due to $Q_{HG}$ and $Q_{3G}$ are smaller by more than a factor of~${\cal O} (20)$ compared to the contributions from $Q_{bG}$.   Missing higher-order QCD effects in~(\ref{eq:GammaSMEFTN3LO}) can  therefore only have a relative numerical impact of  a  few permille once existing experimental limits on the Wilson coefficients of the operators in~(\ref{eq:operators}) are taken into account.

\subsection[Differential $pp \to Zh \to \ell^+ \ell^- b \bar b$  cross section in the SMEFT]{Differential $\bm{pp \to Zh \to \ell^+ \ell^- b \bar b}$  cross section in the SMEFT}

In our differential analysis we select events with two charged leptons (electrons or muons) to explore the $Zh \to \ell^+ \ell^- b \bar b$ signature. The leptons are required to have a transverse momentum of $p_{T,\ell} > 15 \, {\rm GeV}$ and a pseudorapidity of $|\eta_\ell| < 2.5$. The invariant mass of the dilepton pair is restricted to $ m_{\ell^+ \ell^-} \in [75, 105] \, {\rm GeV}$. The events are furthermore required to have at least two $b$-jets, which are reconstructed using the anti-$k_t$ algorithm~\cite{Cacciari:2008gp} as implemented in {\tt FastJet}~\cite{Cacciari:2011ma}. We impose  transverse momentum cuts of $p_{T, b} > 25 \, {\rm GeV}$ and a rapidity threshold of $|\eta_b| < 2.5$ on the $b$-jets. The definition of potential additional jets use the same thresholds as  those of the $b$-jets. The~dominant background processes are $Z + {\rm jets}$, $t \bar t$, single-top and diboson production. The latter three types of backgrounds can be substantially reduced by requiring large values of~$p_{T,Z}$~\cite{Butterworth:2008iy}. Hence, to improve the signal-to-background ratio we impose $p_{T,Z} \in [150, 250] \, {\rm GeV}$. Notice that this $p_{T,Z}$ requirement corresponds to the second resolved $p_{T,Z}$ bin as recommended in the stage~1.2 simplified template cross sections~(STXS) framework~\cite{Andersen:2016qtm,Berger:2019wnu,Amoroso:2020lgh} which is also implemented in the latest ATLAS LHC~Run~II~measurements of the $pp \to Zh \to\ell^+ \ell^- b \bar b$ process~\cite{ATLAS:2019yhn,ATLAS:2020fcp}. We will also comment on how our results are modified if the other two resolved regions,~i.e.~$p_{T,Z} \in [75, 150] \, {\rm GeV}$ and $ p_{T,Z} > 250 \, {\rm GeV}$, are considered.
 
\begin{figure}[t!]
\begin{center}
\includegraphics[width=0.45\textwidth]{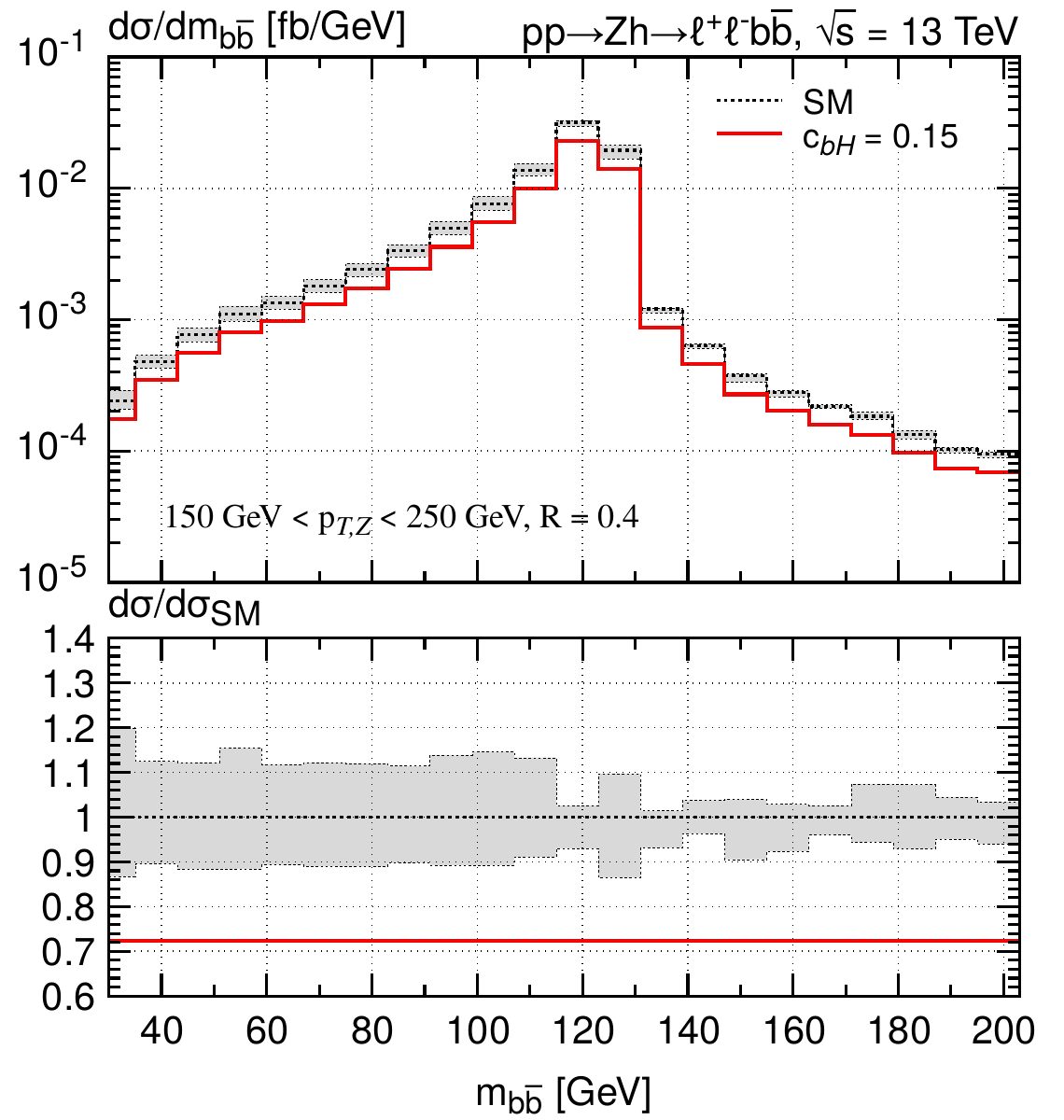} \qquad 
\includegraphics[width=0.45\textwidth]{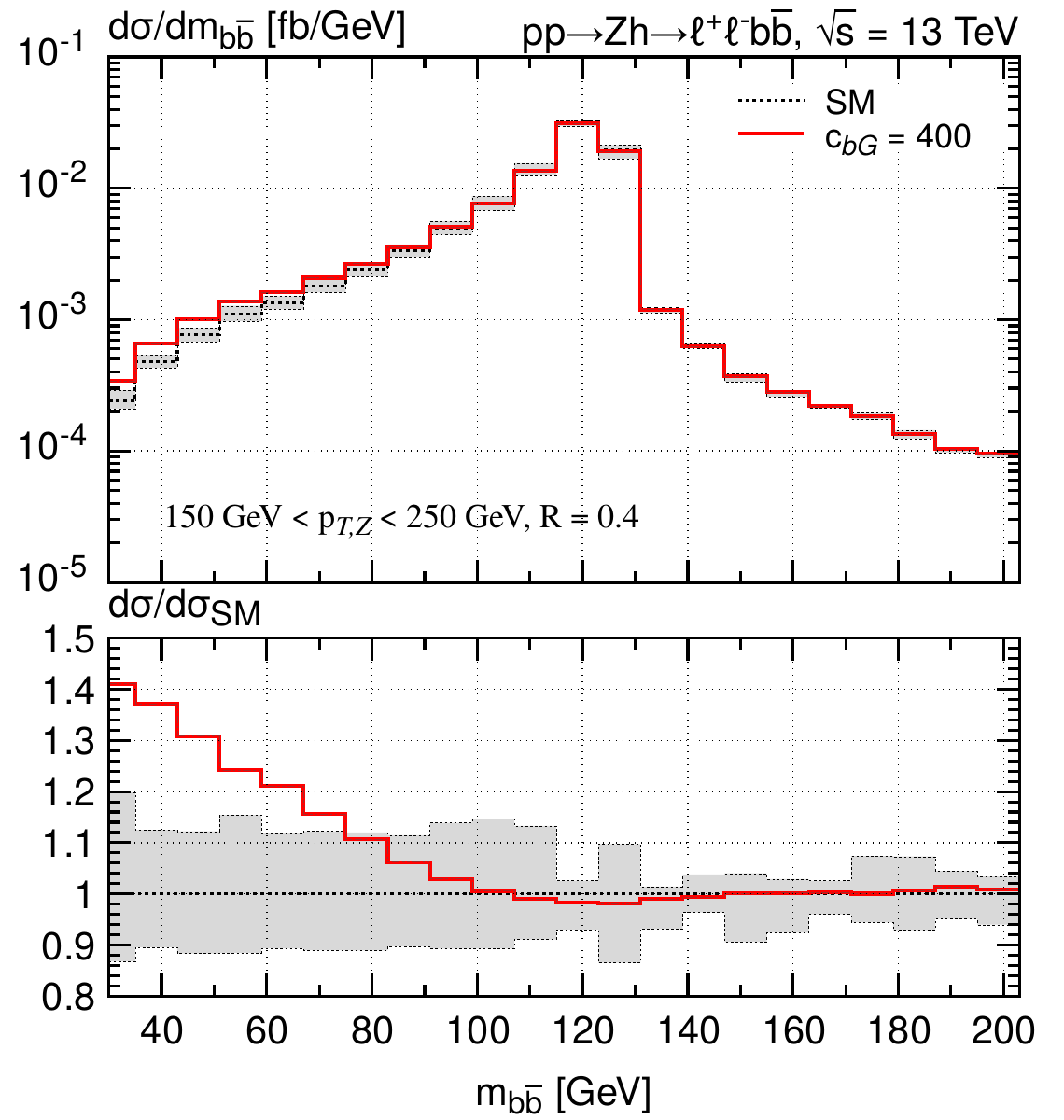}
\end{center}
\vspace{0mm} 
\caption{\label{fig:mbbplots1}  Invariant mass of the two $b$-jets using the anti-$k_t$ algorithm with radius parameter~$R = 0.4$. The red histogram in the left (right) panel corresponds to the prediction for $c_{bH} = 0.15$~($c_{bG} = 400$). For comparison our SM prediction with its scale uncertainty band is shown in black and gray. All results correspond to proton-proton~($pp$) collisions at $\sqrt{s} = 13 \, {\rm TeV}$ and are subject to the fiducial cuts discussed in the main text. The lower panels depict the ratios between the BSM and the SM distributions.}
\end{figure}

The two panels in Figure~\ref{fig:mbbplots1} display  our predictions for the $pp \to Zh \to \ell^+ \ell^- b \bar b$ cross section  differential in the invariant mass of the two $b$-jets, employing a jet radius of $R = 0.4$ in the anti-$k_t$ clustering. If more than two $b$-jets are present the observable $m_{b \bar b}$ is  defined as the invariant mass of the pair of $b$-jets closest to $m_h$.  The black curves correspond to our SM~\NNLOplusPS prediction for the $13 \, {\rm TeV}$ LHC with central renormalisation scale~$\mu_R$ and factorisation scale~$\mu_F$  set according to the \MiNNLOPS~procedure~\cite{Monni:2019whf,Monni:2020nks} and the gray band represents the corresponding perturbative uncertainties. These uncertainties have been obtained from seven-point scale variations enforcing the constraint $1/2 \leq \mu_R/\mu_F \leq 2$ and keeping the scale variation in production and decay correlated. The~same way of estimating perturbative uncertainties is applied to all kinematic distributions that are provided in this section. The red histogram in the left and right panel of  Figure~\ref{fig:mbbplots1} corresponds to the results for $c_{bH} = 0.15$  and $c_{bG} = 400$, respectively.  These values are within the range allowed by~(\ref{eq:cbH95CL}) and (\ref{eq:cbG95CL}). All other Wilson coefficients not specified in a given plot are set to zero.

From the left plot in  Figure~\ref{fig:mbbplots1} it is evident that BSM effects in the form of a non-zero Wilson coefficient~$c_{bH}$ just lead to a rescaling of the~$m_{b \bar b}$ spectrum. This is expected because~$c_{bH}$ is part of the factorisable corrections in~(\ref{eq:cfac}) that just rescales all kinematic distributions by an overall factor.  On the other hand, a non-zero Wilson coefficient~$c_{bG}$ is more interesting, since $c_{bG} \neq 0$ alters the shape of the~$m_{b \bar b}$ distribution with respect to the SM prediction. This can be seen in the right panel of Figure~\ref{fig:mbbplots1}. In fact, one observes that for the choices $c_{bG} = 400$ and $R = 0.4$ the~$m_{b \bar b}$ spectrum receives relative corrections of up to $40\%$ for invariant masses $m_{b \bar b} \simeq 50 \, {\rm GeV}$. The reason for this somewhat surprising feature is the structure of the tree-level squared matrix element, given in~(\ref{eq:MEybdgLO}), that modifies the $h \to b \bar b g$ process and constitutes the leading~$Q_{bG}$ contribution. The~corresponding  Feynman diagram is shown in~Figure~\ref{fig:diagramsQbG} in  the upper row on the left-hand side. From~(\ref{eq:MEybdgLO}) one observes that the probability for emitting a gluon is flat in phase space. In contrast, the real emission contribution to the differential decay rate $h \to b \bar b g$ in the SM is divergent when the radiated gluon becomes unresolved, i.e.~soft or collinear to one of the bottom quarks, and therefore such emissions are favoured. As a result, configurations where the total invariant mass~$m_{b\bar{b}g}=m_h$ of the $b \bar b g$ system  is  shared equally  between the three individual partons occur much more frequently in the former  than in the latter  case, where the bottom quarks typically carry most of the energy which leads to an invariant mass distribution that is strongly peaked at~$m_{b \bar b} \simeq m_h$. We add that changing the sign of  $c_{bH}$ or $c_{bG}$ will also change the sign of the relative corrections due to the considered SMEFT~operators. 

\begin{figure}[t!]
\begin{center}
\includegraphics[width=0.45\textwidth]{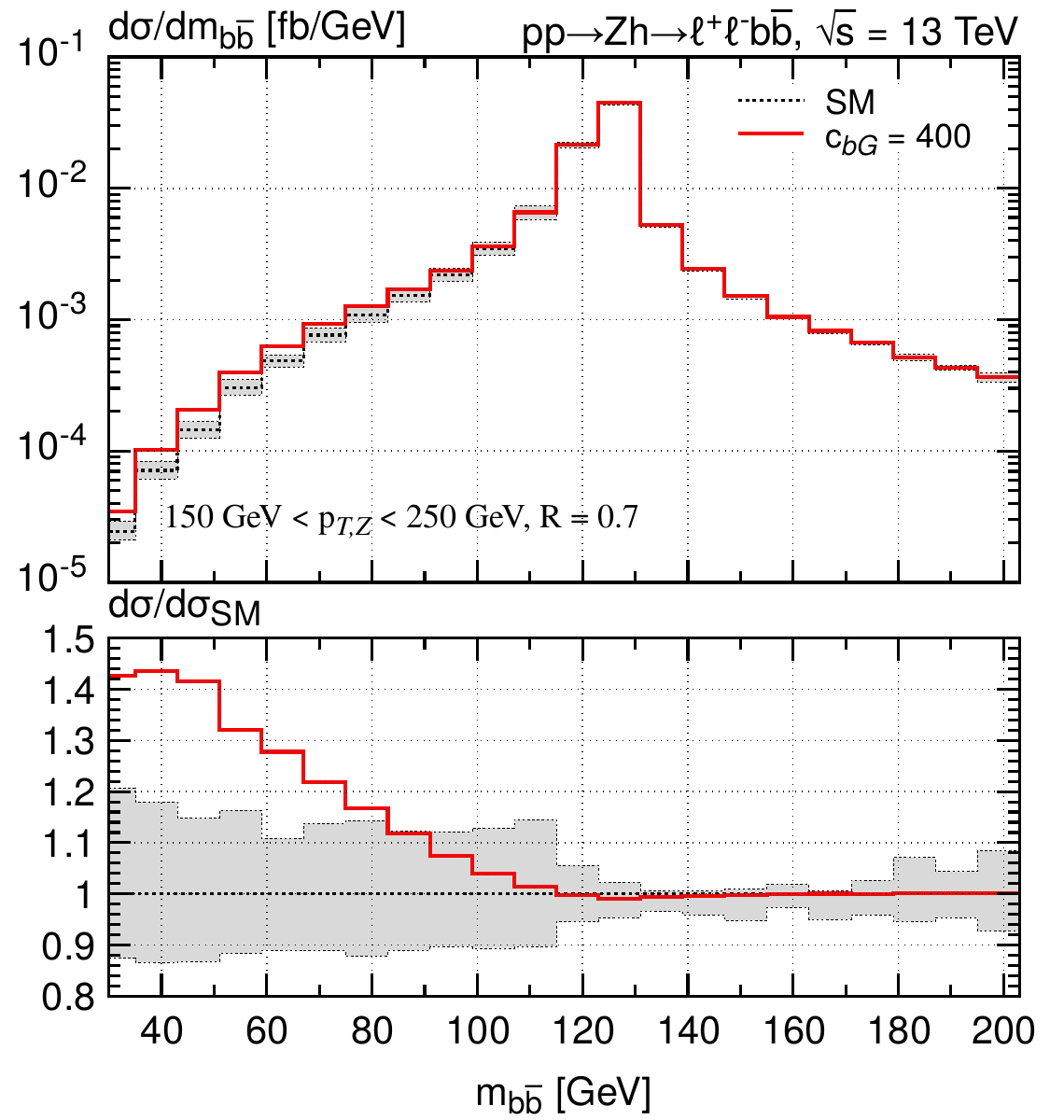} \qquad 
\includegraphics[width=0.45\textwidth]{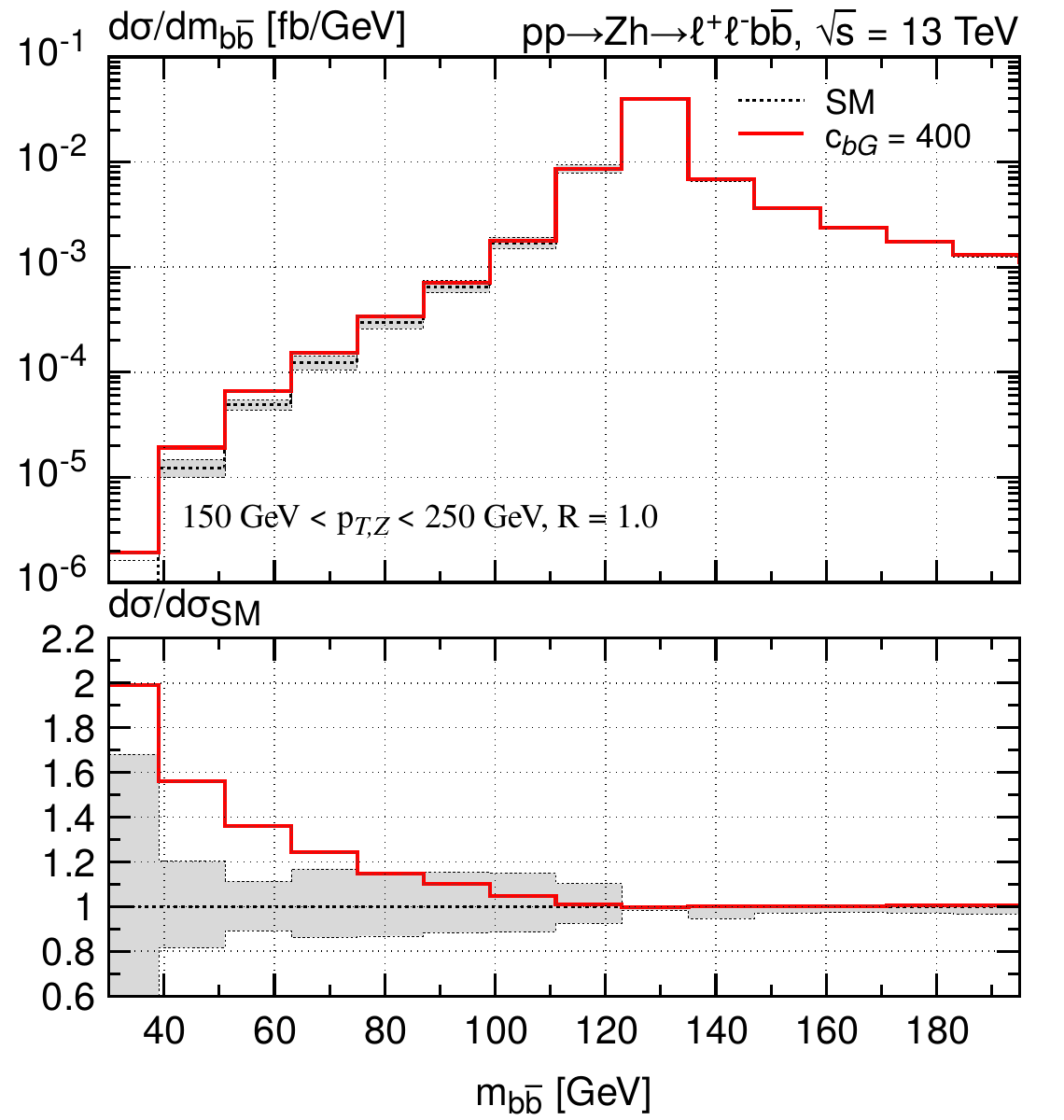}
\end{center}
\vspace{0mm} 
\caption{\label{fig:mbbplots2}  Same as the right panel in Figure~\ref{fig:mbbplots1}, but for~$R = 0.7$~(left) and $R=1.0$~(right). See the main text for further details. } 
\end{figure}

Notice that in the case of $c_{bG} \neq 0$ the shape of the $m_{b \bar b}$ distribution depends on the jet radius $R$ used to identify $b$-jets. To illustrate this feature we display in Figure~\ref{fig:mbbplots2} two additional spectra assuming again $c_{bG} = 400$, but taking~$R = 0.7$~and $R=1.0$ instead of the standard choice $R = 0.4$. One observes that  the corrections due to~$Q_{bG}$ are on average pushed towards lower values of $m_{b \bar b}$ when the jet radius $R$ is increased. We further add in this context that at ${\cal O} (\alpha_s^3)$ insertions of the operator~$Q_{bG}$ lead to a one-loop contribution to the $h \to b \bar b g$ amplitude,  tree-level contributions  to the $h \to b \bar b q \bar q$ and  $h \to b \bar b g g$ processes and two-loop effects in the $h \to b \bar b$ amplitude (cf.~Figure~\ref{fig:diagramsQbG}). While~the two-loop corrections contribute only at $m_{b \bar b} = m_h$, the other two types of contributions are again spread over the phase space. Since the~${\cal O} (\alpha_s^3)$ non-factorisable corrections due to~$Q_{bG}$ are relatively large in the case of the inclusive $h \to b \bar b$ decay width in~(\ref{eq:GammaSMEFTN3LO}), including them in the calculation of the differential cross sections for the full process $pp \to Zh \to \ell^+ \ell^- b \bar b$  is necessary if one wants to describe kinematic distributions such as $m_{b \bar b}$ accurately. We~also note that, while the results shown in Figures~\ref{fig:mbbplots1} and~\ref{fig:mbbplots2} have been obtained for $p_{T,Z} \in [150, 250] \, {\rm GeV}$, qualitatively similar modifications of the $m_{b \bar b}$ distribution due to~$Q_{bG}$ are found when the transverse momentum of the $Z$ boson is restricted to the other two  stage~1.2 STXS  regions with $p_{T,Z} \in [75, 150] \, {\rm GeV}$ and $ p_{T,Z} > 250 \, {\rm GeV}$. As for the shapes, the observed differences depend on the exact $b$-jet definition. In fact, they turn out to be more pronounced for a larger radius $R$. 

\begin{figure}[t!]
\begin{center}
\includegraphics[width=0.45\textwidth]{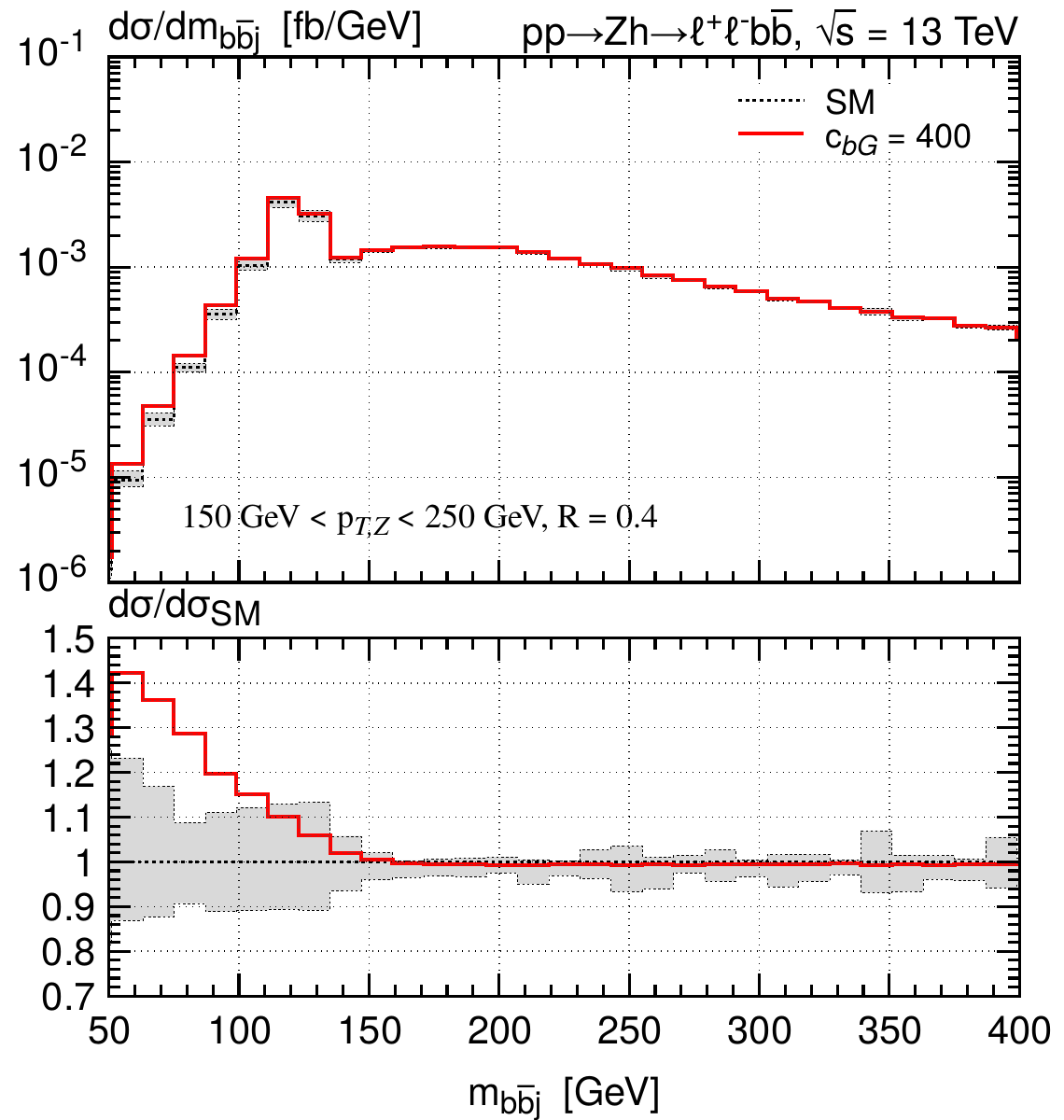} \qquad 
\includegraphics[width=0.45\textwidth]{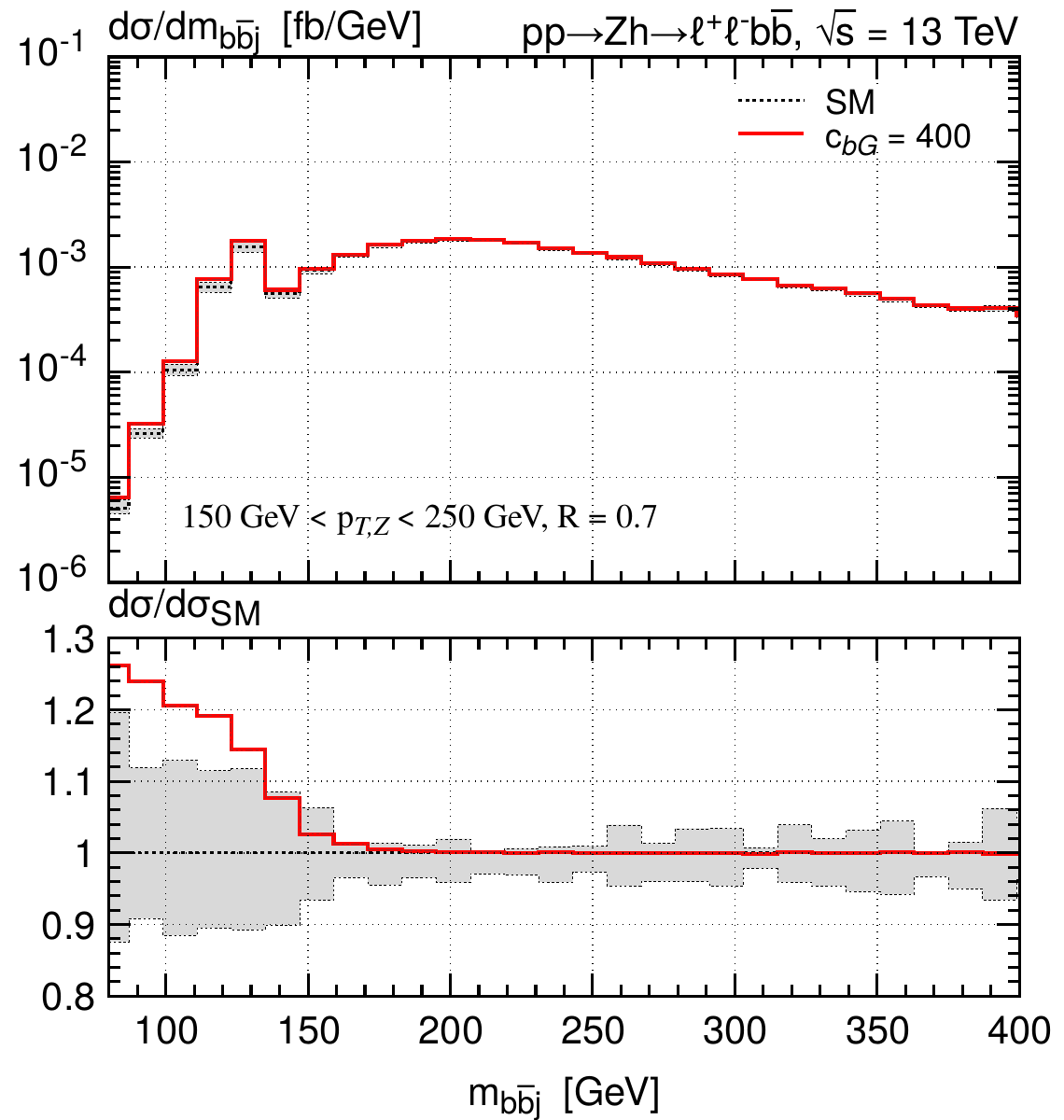}
\end{center}
\vspace{0mm} 
\caption{\label{fig:mbbjplots}  Same as  Figure~\ref{fig:mbbplots2}, but for $m_{b \bar b j}$ using~$R = 0.4$~(left panel) and $R=0.7$~(right panel). See the main text for further details.} 
\end{figure}

Another distribution that features interesting shape changes  in the presence of a non-zero Wilson coefficient~$c_{bG}$ is the invariant  mass $m_{b \bar b j}$ of the two $b$-jets and an extra jet. We built this observable from the set of two $b$-jets and one additional jet whose three-jet invariant mass lies closest  to the Higgs boson mass. Figure~\ref{fig:mbbjplots}~shows two $m_{b \bar b j}$ spectra for $c_{bG} = 400$ with a jet radius of $R=0.4$ and $R = 0.7$ in the left and the right panel, respectively. Also the modifications in the $m_{b \bar b j}$ spectrum due to $c_{bG} \neq 0$ are non-trivial and $R$-dependent. In fact, the relative SMEFT effects are more  pronounced at lower values of $m_{b \bar b j}$ and can reach up to around 40\%. This feature can be qualitatively understood by remembering that the leading $Q_{bg}$  contributions arise from the process $h \to b \bar b g$, see~(\ref{eq:MEybdgLO}), while in the SM the corresponding matrix element is part of the NLO corrections. Relative to the inclusive $h \to b \bar b$ decay width, events with an additional jet will therefore occur more likely in SMEFT scenarios with $c_{bG} \neq 0$ than in the SM. Moreover, since the largest corrections in the invariant mass distributions of the two $b$-jets arise at $m_{b \bar b} \simeq 40 \, {\rm GeV}$ (cf.~the right panel in~Figure~\ref{fig:mbbplots1} and the two panels in~Figure~\ref{fig:mbbplots2}) and we impose $p_{T,j} > 25 \, {\rm GeV}$, one expects to see an excess of events at $m_{b \bar b j} \simeq 60 \, {\rm GeV}$. Indeed, this is  what is observed in the left panel of Figure~\ref{fig:mbbjplots}. It is also clearly visible from the two plots in Figure~\ref{fig:mbbjplots} that increasing the jet radius from $R=0.4$ to $R=0.7$ will result in a migration of events to higher  $m_{b\bar bj}$ values, since a larger jet radius will collect more radiation, leading on average to a larger three-jet invariant mass. We finally mention that in the recent ATLAS analysis~\cite{ATLAS:2020fcp} of $pp \to Vh$ production with $h \to b \bar b$ decay the mass $m_{b \bar b j}$ of the three-jet system is already used as an input to build the multivariate discriminant in the case of three-jet events (see~Table~5 of that publication). We believe that besides $m_{b \bar b}$ the variable $m_{b \bar b j}$ can play an important role in the context of multivariate discriminants tailored to  put constraints on the Wilson coefficient of the operator $Q_{bG}$. 

Let us add that apart from the $m_{b \bar b}$ and $m_{ b \bar b j}$ spectra we have identified additional kinematic distributions that are sensitive to the non-factorisable corrections resulting from the operator~$Q_{bG}$. For instance, also the transverse momentum $p_{T,Z}$ of the $Z$ boson is modified in a non-trivial fashion by~$c_{bG} \neq 0$. However, in view of~(\ref{eq:max}) the effects in~$p_{T,Z}$  cannot exceed the percent level, and therefore this variable taken by itself will have only a rather limited constraining power at the LHC. Similar statements apply to the transverse momentum~$p_{T,b \bar b}$ of the $b$-jet pair. Incorporating the latter observables into a multivariate discriminant may however enhance the overall sensitivity to BSM effects associated to $c_{bG} \neq 0$. An analysis of this issue is clearly beyond the scope of this article. Likewise we also do not attempt to derive bounds on the Wilson~coefficients~$c_{bH}$ and $c_{bG}$ using existing~\cite{ATLAS:2018kot,CMS:2018nsn,ATLAS:2019yhn,ATLAS:2020fcp} or hypothetical~\cite{ATLAS:2018jlh,CMS:2018qgz} differential~LHC~data, leaving such an exercise  for future research. 

\section{Conclusions}
\label{sec:conclusions}

In this article, we have presented novel predictions for the $pp \to Zh \to \ell^+ \ell^- b \bar b$  process within the SMEFT. NNLO QCD corrections for both $pp \to Zh$ production and $h \to b \bar b$ decay have been calculated for a subset of six dimension-six operators appearing in the full SMEFT Lagrangian.  These~fixed-order predictions have been consistently matched to a PS, allowing for a realistic exclusive description of the process at the level of hadronic events while retaining NNLO~QCD accuracy. All SMEFT corrections that are associated to the considered operators and can exceed a percent at the level of the total rate after taking into account the existing experimental bounds on the relevant Wilson coefficients have been implemented into a dedicated MC generator.

Using our \NNLOplusPS~generator we have performed a phenomenological study of the  impact of SMEFT contributions on  several kinematic distributions in $pp \to Zh \to \ell^+ \ell^- b \bar b$  production considering simple benchmark scenarios.  First, we have employed our calculation to extend the  SMEFT computation of the inclusive  $h \to b \bar b$  decay width~\cite{Gauld:2016kuu} to~${\cal O} (\alpha_s^3)$ in the five-flavour scheme.  Our~result for the N$^3$LO corrections due to the chromomagnetic dipole-type operator~$Q_{bG}$ is new. These higher-order QCD effects enhance the  numerical impact of~$Q_{bG}$ by about~60\% compared to the~${\cal O} (\alpha_s^2)$ contributions that were known before. In our phenomenological analysis of differential distributions in $pp \to Zh \to \ell^+ \ell^- b \bar b$ production we have considered explicitly  the~$m_{b \bar b}$ and~$m_{b \bar b j}$ spectra and illustrated how SMEFT effects can change both the normalisation and the shape of these distributions. While~the factorisable corrections associated to the operators $Q_{H \Box}$, $Q_{HD}$ and $Q_{bH}$ act  as an overall rescaling factor on all kinematic observables, the non-factorisable contributions that arise from the operator~$Q_{bG}$ in~(\ref{eq:operators}) lead to non-trivial modifications on the shapes of certain distributions. By studying the dependence on the jet radius $R$  that enters the anti-$k_t$ jet clustering algorithm, we have shown that these shape changes depend on the $b$-jet definition, a feature that to our knowledge has not been discussed in the SMEFT context before.  It might be possible to exploit the observed $R$-dependence of the kinematic distributions such as~$m_{b \bar b}$ or~$m_{b \bar b j}$ to enhance the LHC sensitivity to the operator~$Q_{bG}$ that is at present only very weakly constrained. Furthermore, combining the invariant masses~$m_{b \bar b}$ and~$m_{b \bar b j}$ with other observables such as $p_{T,Z}$ and $p_{T, b \bar b}$ in multivariate discriminants that describe the full kinematics of the selected events is likely to enhance the sensitivity of the LHC to BSM realisations with a non-zero Wilson coefficient $c_{bG}$.\footnote{At future $e^+ e^-$ colliders it might be possible to exploit global event-shape variables in the $h \to b \bar b$ decay~\cite{Coloretti:2022jcl} to put constraints on the Wilson coefficient~$c_{bG}$.} Since~we believe that our MC generator  should prove useful for everyone interested in comparing LHC data to  high-precision SMEFT predictions, we will make the relevant codes to simulate \NNLOplusPS~events  for the $pp \to Zh \to \ell^+ \ell^- b \bar b$ process in the SMEFT publicly available within the {\tt POWHEG-BOX}~framework~\cite{POWHEGBOXRES}. 

Our analysis of SMEFT effects in the  $pp \to Zh \to \ell^+ \ell^- b \bar b$  process includes only a limited subset of dimension-six operators. For instance effective interactions that modify the couplings of the Higgs to EW gauge bosons are not considered in this work. Since it is known~\cite{Mimasu:2015nqa,Degrande:2016dqg,Alioli:2018ljm,Bizon:2021rww,Maltoni:2013sma,Greljo:2017spw} that the latter type of SMEFT contributions can lead to phenomenological relevant effects in $Zh$ production, we plan  to apply the discussed \NNLOplusPS~method to non-trivial insertions of purely EW operators as well. Since the leading EW operators contributing to $Zh$ production play only a minor role in the $h \to b \bar b$ decay and vice versa, obtaining  \NNLOplusPS~accuracy for a suitable enlarged set of  dimension-six operators in the SMEFT  seems possible following the methodology developed here. 

\acknowledgments  We are grateful to Rhorry Gauld for helpful discussions and useful comments that helped us to improve the manuscript. We also thank Marumi Kado for interesting conversations. The simulations related to this work  have used the Max Planck Computing and Data Facility (MPCDF) in Garching. Silvia Zanoli is supported by the International Max Planck Research School (IMPRS) on “Elementary Particle Physics”. 

\begin{appendix}

\section{Squared matrix elements}
\label{app:MEs}

Below  we provide the expressions for the squared matrix elements for the non-factorisable SMEFT terms of~${\cal O} (y_b^2  \hspace{0.25mm} C_{bG})$ to the $h \to b \bar b$ decay distributions that are included in our  \NNLOplusPS~generator. All matrix elements are computed using conventional dimensional regularisation for both UV and~IR singularities in $d = 4 -2 \hspace{0.125mm} \epsilon$ dimensions with $\overline{\rm MS}$ subtraction. The~actual generation and computation of squared matrix elements relies on the {\tt Mathematica} packages {\tt FeynRules}~\cite{Alloul:2013bka}, {\tt FeynArts}~\cite{Hahn:2000kx}, {\tt FormCalc}~\cite{Hahn:1998yk,Hahn:2016ebn}, {\tt Package-X}~\cite{Patel:2015tea} and {\tt LiteRed}~\cite{Lee:2013mka} as~well~as~{\tt VSM}. 

The interference term between the tree-level matrix elements of $h (p_h) \to b (p_1) \bar b (p_2)  g (p_3)$ with an insertion of $Q_{bG}$ and the corresponding SM contribution gives rise to a contribution of~${\cal O} (y_b^2  \hspace{0.25mm} \alpha_s^2  \hspace{0.25mm} C_{bG})$. It   takes the following form 
 \beq \label{eq:MEybdgLO}
{\cal S}_{b \bar b g}^{ (0)} = 2 \hspace{0.25mm} {\rm Re}  \, \Big \langle {\cal M}_{y_b, b \bar b g}^{(0)} \, \Big | \, {\cal M}_{C_{bG}, b \bar b g}^{(0)} \Big  \rangle = 8 \hspace{0.25mm} \alpha_s^2 \hspace{0.25mm} C_F \hspace{0.25mm} C_A \hspace{0.25mm} m_h^2 \hspace{0.5mm} y_b^2 \hspace{0.75mm} \frac{{\rm Re} \left ( C_{bG} \right)}{\Lambda^2}  \,.
\eeq
Here~${\cal M}_{y_b, b \bar b g}^{(0)}$ and~${\cal M}_{C_{bG}, b \bar b g}^{(0)}$ denote the SM and SMEFT tree-level $h \to b \bar b g$ matrix element, respectively,  and $C_F = 4/3$ and $C_A = 3$ are the relevant colour factors. 

At~${\cal O} (y_b^2  \hspace{0.25mm} \alpha_s^3  \hspace{0.25mm} C_{bG})$ the interference term between the $Q_{bG}$ and the SM contribution to $h (p_h) \to b (p_1) \bar b (p_2)  g (p_3)$ can be written as follows 
\beq  \label{eq:MEybdgNLO}
{\cal S}_{b \bar b g}^{(1)}  = 2 \, {\rm Re}  \, \Big \langle {\cal M}_{y_b, b \bar b g}^{(1)} \, \Big | \, {\cal M}_{C_{bG}, b \bar b g}^{(0)} \Big  \rangle +
2 \, {\rm Re}  \, \Big \langle {\cal M}_{y_b, b \bar b g}^{(0)} \, \Big | \, {\cal M}_{C_{bG}, b \bar b g}^{(1)} \Big  \rangle \,,
\eeq 
where~${\cal M}_{y_b, b \bar b g}^{(1)}$ and~${\cal M}_{C_{bG}, b \bar b g}^{(1)}$ denote the SM and SMEFT one-loop $h \to b \bar b g$ matrix element. To write our result for~${\cal S}_{b \bar b g}^{(1)}$ in a compact form we will use 
\beq \label{eq:yij}
y_{ij} = \frac{2 \hspace{0.25mm} p_i \cdot p_j}{m_h^2} \,,
\eeq
i.e.~twice the dot-product of two external four-momenta $p_i$ and $p_j$ divided by the Higgs boson mass squared. For the sum of   terms in~(\ref{eq:MEybdgNLO}) we obtain 
\bea \label{eq:MEybdg10}
\begin{split}
{\cal S}_{b \bar b g}^{(1)}  = \frac{\alpha_s}{2 \pi} \, N_\epsilon  \, {\cal S}_{b \bar b g}^{(0)}  \, \Bigg \{  & \displaystyle -\frac{2 \hspace{0.25mm} C_F + C_A}{\epsilon^2}   -  \frac{1}{\epsilon} \, \big  [ \left ( C_A - 2 \hspace{0.25mm} C_F  \right  ) \, L_{12} - C_A  \left ( L_{13} + L_{23} \right ) + \gamma_{g} +  2 \hspace{0.125mm} \gamma_{q} \big  ] \\[2mm]
& \hspace{-2.3cm}  \displaystyle + \left ( C_A-2 \hspace{0.25mm} C_F \right ) \left [ \frac{1}{2}  \hspace{0.5mm} R \hspace{0.25mm} (y_{12}, y_{13} ) + \frac{1}{2} \hspace{0.5mm} R \hspace{0.25mm} (y_{12}, y_{23} ) + \frac{L_{12}^2}{2}  - L_{12} \right ] \\[2mm]
& \hspace{-2.3cm}  \displaystyle - C_A  \left [ \frac{1}{2} \hspace{0.5mm} R \hspace{0.25mm} (y_{13}, y_{23} ) + \frac{L_{13}^2}{2} + \frac{L_{23}^2}{2}  \right ] -  C_F  \left ( 1 - L_{13} - L_{23} \right )  \\[2mm]
& \hspace{-2.3cm}  \displaystyle + 3 \hspace{0.5mm} \big ( 2 \hspace{0.25mm} C_F  + C_A \big  ) \hspace{0.5mm} \zeta_2 +  \frac{1}{2} \left (C_A - C_F \right ) \left ( y_{13} + y_{23} \right )  +  \gamma_{bG}  + \left ( \gamma_g + 2 \hspace{0.125mm} \gamma_q +  \gamma_{bG} \right )  L \hspace{0.5mm}  \Bigg \} \, .
\end{split}
\eea
Here we have defined 
\beq \label{eq:Neps}
N_\epsilon =  \frac{e^{\epsilon \hspace{0.25mm} \gamma_E} }{\Gamma  \hspace{0.25mm}  ( 1 - \epsilon )}  \hspace{0.25mm} \left  ( \frac{\mu^2}{m_h^2} \right )^\epsilon  = 1 + L \hspace{0.25mm} \epsilon + \left [ L^2 - \zeta_2 \right ] \frac{\epsilon^2}{2} \,,
\eeq
where $\Gamma  \hspace{0.25mm}  (z)$ is the gamma function, $\gamma_E \simeq 0.577$  is the Euler-Mascheroni constant, $\zeta_2 = \pi^2/6$ is the Riemann Zeta function of two and 
\beq \label{eq:LijL}
L_{ij} = \ln y_{ij} \,, \qquad  L = \ln \frac{\mu^2}{m_h^2} \,.
\eeq
We have furthermore introduced 
\beq \label{eq:Rxy}
R (x, y) = {\rm Li}_2 \hspace{0.25mm} ( 1 - x )  + {\rm Li}_2  \hspace{0.25mm}  ( 1 - y )  + \ln x \, \ln y - \zeta_2 \,, 
\eeq
with~${\rm Li}_2(z)$  the dilogarithm. The anomalous dimensions $\gamma_g$ and $\gamma_q$ appearing in~(\ref{eq:MEybdg10}) can for example be found in the classic work~\cite{Catani:1998bh} on the IR structure of QCD scattering amplitudes, while calculations of the    anomalous dimension of the  operator $Q_{bG}$ have been performed in~\cite{Misiak:1994zw,Gorbahn:2005sa}. The needed anomalous dimensions read 
\beq \label{eq:ADMs}
\gamma_{g} = \frac{11 \hspace{0.25mm} C_A}{6} - \frac{N_f}{3} \,, \qquad   \gamma_q = \frac{3 \hspace{0.25mm} C_F}{2} \,, \qquad \gamma_{bG} = 2 \hspace{0.25mm} C_A  - 8  \hspace{0.25mm}C_F \,,
\eeq
where $N_f$ denotes the number of active quark flavours. We add that as a cross-check of our SMEFT calculation, we have also computed the interference between the tree-level and one-loop $h \to b \bar b g$ matrix elements in the SM. Our SM calculation reproduces the expression given in~(A.6) of the publication~\cite{DelDuca:2015zqa}.\footnote{The terms $\ln y_{ij}^2$  in~(A.6)  of~\cite{DelDuca:2015zqa} are obvious misprints that should in fact read $\ln^2 y_{ij}$.} 

We also need the~${\cal O} (y_b^2  \hspace{0.25mm} \alpha_s^3  \hspace{0.25mm} C_{bG})$  interferences for all relevant Higgs to four parton scattering processes. To write these squared tree-level amplitudes in a compact form  we introduce
\beq \label{eq:yijk}
y_{ijk} = y_{ij} + y_{ik} + y_{jk} \,. 
\eeq
For $h (p_h) \to b (p_1) \bar b(p_2) q (p_3) \bar q (p_4)$, we obtain 
\beq \label{eq:Sbbqq}
\begin{split}
{\cal S}_{b \bar b q \bar q}^{(0)}  = 2 \hspace{0.25mm} {\rm Re}  \, \Big \langle {\cal M}_{y_b, b \bar b q \bar q}^{(0)} \, \Big | \, {\cal M}_{C_{bG}, b \bar b q \bar q}^{(0)} \Big  \rangle 
= 2 \hspace{0.25mm} \pi \hspace{0.25mm}  \alpha_s  \hspace{0.5mm} {\cal S}_{b \bar b g}^{(0)}  \hspace{0.5mm}  \frac{1}{m_h^2} \hspace{0.5mm}  g_{b \bar b q \bar q} \left (p_1,p_2,p_3,p_4 \right ) \,,
\end{split} 
\eeq
where 
\bea \label{eq:gbbqq}
\begin{split}
g_{b \bar b q \bar q} \left (p_1,p_2,p_3,p_4 \right )  & =  -\frac{y_{13}}{y_{234}} + \frac{y_{13}}{y_{34} y_{134}}-\frac{2 y_{24} y_{13}}{y_{134} y_{234}}-\frac{\left(y_{14} y_{23}+3 y_{13} y_{24}\right) y_{13}}{y_{34} y_{134} y_{234}} \\[2mm]
& \hspace{-1cm} -\frac{y_{24} \left(y_{13} y_{24}-y_{14} y_{23}\right) y_{13}}{y_{34}^2 y_{134} y_{234}}+ ( 1 \leftrightarrow 2 )  + ( 3 \leftrightarrow 4 ) + ( 1 \leftrightarrow 2, 3 \leftrightarrow 4  )  \,. \hspace{8mm} 
\end{split}
\eea
 
In the case of $h(p_h)  \to b(p_1) \bar b(p_2) b(p_3) \bar b (p_4)$, we find 
\beq
\begin{split}
{\cal S}_{b \bar b b \bar b}^{(0)} & = 2 \hspace{0.25mm} {\rm Re}  \, \Big \langle {\cal M}_{y_b, b \bar b b \bar b}^{(0)} \, \Big | \, {\cal M}_{C_{bG}, b \bar b b \bar b}^{(0)} \Big  \rangle   \\[2mm]
& = 2 \hspace{0.25mm} \pi \hspace{0.25mm}  \alpha_s  \hspace{0.5mm} {\cal S}_{b \bar b g}^{(0)}  \hspace{0.5mm}  \frac{1}{m_h^2} \, \bigg [   \big( C_A - 2 \hspace{0.25mm}C_F \big )  \hspace{0.5mm} f_{b \bar b b \bar b} \left (p_1,p_2,p_3,p_4 \right ) + g_{b \bar b b \bar b} \left (p_1,p_2,p_3,p_4 \right ) \bigg ]  \,,
\end{split}
\eeq
where 
\begin{align}
 f_{b \bar b b \bar b} \left (p_1,p_2,p_3,p_4 \right )  & = 
 \frac{y_{12}^5}{y_{14} y_{123} y_{124} y_{134} y_{234}}+\frac{y_{12}^5}{y_{23} y_{123} y_{124} y_{134} y_{234}}+\frac{\left(4 y_{13}+y_{23}+2
   y_{24}-1\right) y_{12}^4}{y_{14} y_{123} y_{124} y_{134} y_{234}} \nonumber \\[2mm] 
& \hspace{-3cm} +\frac{\left(y_{13}+y_{14}+4 y_{24}-1\right) y_{12}^4}{y_{23} y_{123} y_{124} y_{134}
   y_{234}}+\frac{y_{13} \left(y_{23}+3 y_{24}-4\right) y_{12}^3}{y_{14} y_{123} y_{124} y_{134} y_{234}}-\frac{\left(2 y_{13}+y_{14}+y_{23}+y_{24}+2
   y_{34}\right) y_{12}^3}{y_{123} y_{124} y_{134} y_{234}}  \nonumber \\[2mm]
& \hspace{-3cm} +\frac{\left(y_{24} \left(2 y_{13}+y_{14}+6 y_{24}-4\right)-y_{14} y_{34}\right) y_{12}^3}{y_{23}
   y_{123} y_{124} y_{134} y_{234}}+\frac{\left(y_{24} \left(y_{23}+y_{24}-1\right)-y_{23} y_{34}\right) y_{12}^3}{y_{14} y_{123} y_{124} y_{134}
   y_{234}}  \nonumber \\[2mm]
& \hspace{-3cm} +\frac{6 y_{13}^2 y_{12}^3}{y_{14} y_{123} y_{124} y_{134} y_{234}}+\frac{y_{13} y_{24}^2 y_{12}^3}{y_{23} y_{34} y_{123} y_{124} y_{134}  
   y_{234}} +\frac{y_{13}^2 y_{24} y_{12}^3}{y_{14} y_{34} y_{123} y_{124} y_{134} y_{234}}  \nonumber \\[2mm]
& \hspace{-3cm} -\frac{\left(y_{13} y_{23}-2 y_{14} y_{23}+2 y_{13} y_{24}+y_{14}
   y_{24}\right) y_{12}^3}{y_{34} y_{123} y_{124} y_{134} y_{234}}-\frac{y_{13}^2 y_{12}^2}{y_{123} y_{124} y_{134} y_{234}}-\frac{y_{13}^2 y_{24}
   y_{12}^2}{y_{23} y_{123} y_{124} y_{134} y_{234}}  \nonumber \\[2mm]
& \hspace{-3cm} +\frac{y_{13}^2 \left(3 y_{24}-4\right) y_{12}^2}{y_{14} y_{123} y_{124} y_{134} y_{234}}-\frac{y_{24}^2
   y_{34} y_{12}^2}{y_{14} y_{123} y_{124} y_{134} y_{234}} +\frac{\left(4 y_{24}^3-4 y_{24}^2-y_{34} \left(5 y_{14}+y_{34}\right) y_{24}+2 y_{14}
   y_{34}^2\right) y_{12}^2}{y_{23} y_{123} y_{124} y_{134} y_{234}}  \nonumber \\[2mm] 
& \hspace{-3cm} +\frac{y_{13} \left(3 y_{24}^2-4 y_{34} y_{24}-y_{34}+y_{14} \left(y_{34}-5
   y_{24}\right)\right) y_{12}^2}{y_{23} y_{123} y_{124} y_{134} y_{234}}-\frac{y_{13} \left(3 y_{14}+9 y_{23}+4 \left(3 y_{24}+y_{34}\right)\right)
   y_{12}^2}{y_{123} y_{124} y_{134} y_{234}} \nonumber \\[2mm] 
& \hspace{-3cm} -\frac{y_{13} \left(5 y_{23} \left(y_{24}+y_{34}\right)+y_{24} \left(y_{24}+5 y_{34}\right)\right)
   y_{12}^2}{y_{14} y_{123} y_{124} y_{134} y_{234}}+\frac{\left(y_{14}^2-9 y_{24} y_{14}+y_{23}^2-y_{24} \left(3 y_{23}+y_{24}+5 y_{34}\right)\right)
   y_{12}^2}{y_{123} y_{124} y_{134} y_{234}} \nonumber \\[2mm]
& \hspace{-3cm} +\frac{3 y_{13}^3 y_{12}^2}{y_{14} y_{123} y_{124} y_{134} y_{234}}-\frac{2 y_{13}^2 y_{23} y_{12}^2}{y_{34}
   y_{123} y_{124} y_{134} y_{234}}+\frac{y_{13} y_{24}^2 \left(3 \left(y_{13}+y_{14}\right)+2 y_{24}\right) y_{12}^2}{y_{23} y_{34} y_{123} y_{124} y_{134}
   y_{234}} \nonumber \\[2mm]
& \hspace{-3cm} -\frac{y_{13} \left(3 y_{23}^2+5 \left(y_{14}+y_{23}\right) y_{24}\right) y_{12}^2}{y_{34} y_{123} y_{124} y_{134} y_{234}}+\frac{y_{14} \left(-2
   y_{24}^2-3 y_{14} y_{24}+2 y_{23} \left(y_{14}+y_{23}\right)\right) y_{12}^2}{y_{34} y_{123} y_{124} y_{134} y_{234}} \nonumber \\[2mm]
& \hspace{-3cm} +\frac{y_{13}^2 y_{24} \left(2
   y_{13}+3 \left(y_{23}+y_{24}\right)\right) y_{12}^2}{y_{14} y_{34} y_{123} y_{124} y_{134} y_{234}}-\frac{y_{13}^3 y_{12}}{y_{123} y_{124} y_{134}
   y_{234}}+\frac{\left(y_{24}-1\right) y_{24}^3 y_{12}}{y_{23} y_{123} y_{124} y_{134} y_{234}} \nonumber \\[2mm]
& \hspace{-3cm} +\frac{y_{13} y_{24} \left(y_{13}+y_{24}\right) \left(2
   y_{13}+y_{24}\right) y_{12}}{y_{14} y_{123} y_{124} y_{134} y_{234}}-\frac{y_{13} \left(y_{24}^2+4 \left(y_{14}+y_{23}\right) y_{24}+6 y_{14}
   y_{23}\right) y_{12}}{y_{123} y_{124} y_{134} y_{234}} \nonumber \\[2mm]
& \hspace{-3cm} +\frac{3 y_{13} y_{24}^2 \left(y_{24}-y_{34}\right) y_{12}}{y_{23} y_{123} y_{124} y_{134}
   y_{234}}+\frac{3 y_{13}^2 y_{24} \left(y_{24}-y_{34}\right) y_{12}}{y_{23} y_{123} y_{124} y_{134} y_{234}}+\frac{y_{13}^3 \left(y_{24}-y_{34}\right)
   y_{12}}{y_{23} y_{123} y_{124} y_{134} y_{234}} \nonumber \\[2mm]
& \hspace{-3cm} -\frac{y_{13}^2 \left(3 y_{14}+4 y_{23}+y_{24}+4 y_{34}\right) y_{12}}{y_{123} y_{124} y_{134}
   y_{234}}-\frac{y_{24} \left(6 y_{14} y_{23}+3 y_{24} y_{23}+4 y_{14} y_{24}+4 y_{24} y_{34}\right) y_{12}}{y_{123} y_{124} y_{134} y_{234}} \nonumber \\[2mm]
& \hspace{-3cm} -\frac{3
   y_{14} y_{23} y_{24}^2 y_{12}}{y_{34} y_{123} y_{124} y_{134} y_{234}}+\frac{2 y_{13}^3 y_{24}^2 y_{12}}{y_{23} y_{34} y_{123} y_{124} y_{134}
   y_{234}}-\frac{y_{13}^3 \left(y_{23}-3 y_{24}\right) y_{12}}{y_{34} y_{123} y_{124} y_{134} y_{234}} \nonumber \\[2mm]
& \hspace{-3cm} +\frac{y_{13}^4 y_{24} y_{12}}{y_{14} y_{34} y_{123}
   y_{124} y_{134} y_{234}}+\frac{y_{13} y_{24}^3 \left(2 y_{14}+y_{24}\right) y_{12}}{y_{23} y_{34} y_{123} y_{124} y_{134} y_{234}}+\frac{y_{13}^2
   y_{24}^2 \left(3 y_{23}+2 y_{24}\right) y_{12}}{y_{14} y_{34} y_{123} y_{124} y_{134} y_{234}} \nonumber \\[2mm]
& \hspace{-3cm} +\frac{y_{13}^2 y_{24}^2 \left(3 y_{14}+5 y_{24}\right)
   y_{12}}{y_{23} y_{34} y_{123} y_{124} y_{134} y_{234}}+\frac{y_{13}^3 y_{24} \left(2 y_{23}+5 y_{24}\right) y_{12}}{y_{14} y_{34} y_{123} y_{124} y_{134}
   y_{234}}+\frac{y_{13}^2 \left(6 y_{24}^2-2 y_{14} y_{24}-3 y_{14} y_{23}\right) y_{12}}{y_{34} y_{123} y_{124} y_{134} y_{234}} \nonumber \\[2mm]
& \hspace{-3cm} +\frac{y_{13} y_{24}
   \left(y_{24} \left(3 y_{24}-2 y_{23}\right)-8 y_{14} y_{23}\right) y_{12}}{y_{34} y_{123} y_{124} y_{134} y_{234}}+\frac{2 y_{13} y_{24}
   \left(y_{13}^2+y_{24}^2\right)}{y_{123} y_{124} y_{134} y_{234}}+\frac{y_{13}^2 y_{24}^3 \left(y_{13}+y_{24}\right)}{y_{23} y_{34} y_{123} y_{124}
   y_{134} y_{234}} \nonumber \\[2mm] 
& \hspace{-3cm} +\frac{y_{13}^3 y_{24}^2 \left(y_{13}+y_{24}\right)}{y_{14} y_{34} y_{123} y_{124} y_{134} y_{234}}+\frac{y_{13} y_{24}
   \left(y_{13}+y_{24}\right) \left(y_{13}^2+3 y_{24} y_{13}+y_{24}^2\right)}{y_{34} y_{123} y_{124} y_{134} y_{234}} \nonumber \\[2mm]
& \hspace{-3cm}  + ( 1 \leftrightarrow 3 )  + ( 2 \leftrightarrow 4 ) + ( 1 \leftrightarrow 3, 2 \leftrightarrow 4  )  \,.
\end{align}
Furthermore, 
\beq
 g_{b \bar b b \bar b} \left (p_1,p_2,p_3,p_4 \right ) = g_{b \bar b q \bar q} \left (p_1,p_2,p_3,p_4 \right ) + ( 1 \leftrightarrow 3 )  + ( 2 \leftrightarrow 4 ) + ( 1 \leftrightarrow 3, 2 \leftrightarrow 4  )   \,, 
\eeq
with $g_{b \bar b q \bar q} \left (p_1,p_2,p_3,p_4 \right )$ already defined  in~(\ref{eq:gbbqq}).

For the process $h(p_h) \to b(p_1) \bar b (p_2) g(p_3) g(p_4)$, we finally get 
\beq
\begin{split}
{\cal S}_{b \bar b gg}^{(0)} & = 2 \hspace{0.25mm} {\rm Re}  \, \Big \langle {\cal M}_{y_b, b \bar b gg}^{(0)} \, \Big | \, {\cal M}_{C_{bG}, b \bar b gg}^{(0)} \Big  \rangle   \\[2mm]
& =  2 \hspace{0.25mm} \pi \hspace{0.25mm}  \alpha_s  \hspace{0.5mm} {\cal S}_{b \bar b g}^{(0)}  \hspace{0.5mm}  \frac{1}{m_h^2} \hspace{0.5mm}  \Big [ \,  C_A \hspace{0.25mm} f_{b \bar b g  g} \left (p_1,p_2,p_3,p_4 \right )  + C_F \hspace{0.25mm} g_{b \bar b g g} \left (p_1,p_2,p_3,p_4 \right ) \Big ] \,.
\end{split}
\eeq 
Here 
\begin{align}
f_{b \bar b g g} \left (p_1,p_2,p_3,p_4 \right )  & =  \frac{4 y_{24}^2}{y_{23} y_{34} y_{234}}+\frac{y_{13}^2 y_{24}^2}{2 y_{14} y_{23} y_{34} y_{134} y_{234}}+\frac{\left(10 y_{13}+9 y_{34}\right)
   y_{24}}{y_{14} y_{134} y_{234}}  \nonumber \\[2mm]
& \hspace{-3cm} +\frac{\left(4 y_{24}+19 y_{34}-4\right) y_{24}}{y_{23} y_{134} y_{234}}+\frac{\left(4 y_{14}
   \left(y_{23}+2\right)+y_{13} \left(12 y_{24}+7\right)\right) y_{24}}{2 y_{34} y_{134} y_{234}} +\frac{2 y_{13} \left(y_{13} y_{24}-y_{14} y_{23}\right)
   y_{24}}{y_{34}^2 y_{134} y_{234}} \nonumber \\[2mm]
& \hspace{-3cm} -\frac{2 y_{34}^2}{y_{13} y_{14} y_{134}} +\frac{y_{34}^2 \left(y_{23}+y_{24}+y_{34}-1\right)}{y_{13} y_{14} y_{23}
   y_{134}}+\frac{y_{34}^2 \left(y_{23}+3 y_{24}+2 y_{34}\right)}{y_{13} y_{14} y_{134} y_{234}} \nonumber \\[2mm]
& \hspace{-3cm}  +\frac{18 y_{34}+2 y_{24} \left(4 y_{13}+2 y_{23}+2
   y_{24}+2 y_{34}+19\right)-3}{2 y_{134} y_{234}}+\frac{12 y_{14} y_{24}+5 y_{34} \left(2 y_{24}+3 y_{34}-1\right)}{y_{13} y_{134}
   y_{234}} \nonumber \\[2mm]
& \hspace{-3cm} +\frac{2 y_{14} y_{24} \left(2 y_{24}+3 y_{34}-1\right)+ y_{34} \left(3 y_{34}^2+\left(9 y_{24}-2\right) y_{34}+4 \left(y_{24}-1\right)
   y_{24}\right)}{y_{13} y_{23} y_{134} y_{234}} \nonumber \\[2mm]
& \hspace{-3cm}  +\frac{\left(y_{13} y_{24} \left(6 y_{24}+8 y_{34}-1\right)+y_{34} \left(6 y_{24}^2+2 \left(7
   y_{34}-2\right) y_{24}+y_{34} \left(5 y_{34}-3\right)\right)\right)}{2 y_{14} y_{23} y_{134} y_{234}} \nonumber \\[2mm] 
& \hspace{-3cm} + ( 1 \leftrightarrow 2 ) + ( 3 \leftrightarrow 4 ) + ( 1 \leftrightarrow 2, 3 \leftrightarrow 4  ) \,,  
\end{align}
while 
\begin{align}
g_{b \bar b g g} \left (p_1,p_2,p_3,p_4 \right )  & = -\frac{2 \left(y_{23}+y_{24}+y_{34}-1\right) y_{34}^2}{y_{13} y_{14} y_{23} y_{134}}+\frac{4 y_{34}^2}{y_{13} y_{14} y_{134}}-\frac{2 \left(y_{23}+3
   y_{24}+2 y_{34}\right) y_{34}^2}{y_{13} y_{14} y_{134} y_{234}} \nonumber \\[2mm]
& \hspace{-3cm} +\frac{2 y_{24} \left(-4 y_{24}+\left(y_{34}-14\right) y_{34}+4\right)}{y_{23} y_{134}
   y_{234}}+\frac{2 y_{24} \left(y_{34} \left(y_{24}+2 y_{34}-7\right)+y_{13} \left(y_{24}+2 y_{34}-5\right)\right)}{y_{14} y_{134} y_{234}} \nonumber \\[2mm]
& \hspace{-3cm} +\frac{2 \left(3
   y_{24}^2+\left(4 y_{13}+4 y_{14}+3 y_{23}+13 y_{34}-14\right) y_{24}+y_{34} \left(3 y_{34}-8\right)+2\right)}{y_{134} y_{234}} \nonumber \\[2mm]
& \hspace{-3cm} -\frac{2 \left(2 y_{14}
   y_{24} \left(2 y_{24}+3 y_{34}-1\right)+y_{34} \left(3 y_{34}^2+\left(9 y_{24}-2\right) y_{34}+4 \left(y_{24}-1\right) y_{24}\right)\right)}{y_{13}
   y_{23} y_{134} y_{234}} \nonumber \\[2mm]
& \hspace{-3cm} +\frac{2 \left(y_{14} y_{24} \left(y_{23}+y_{34}-8\right)+y_{34} \left(y_{24}^2+\left(2 y_{23}+2 y_{34}-9\right)
   y_{24}+\left(y_{34}-13\right) y_{34}+5\right)\right)}{y_{13} y_{134} y_{234}} \nonumber \\[2mm]
& \hspace{-3cm} -\frac{2 \left(y_{13} y_{24} \left(y_{24}+2 y_{34}\right)+y_{34} \left(2
   y_{24}^2+\left(5 y_{34}-1\right) y_{24}+y_{34} \left(2 y_{34}-1\right)\right)\right)}{y_{14} y_{23} y_{134} y_{234}} \nonumber \\[2mm]
& \hspace{-3cm} + ( 1 \leftrightarrow 2 ) + ( 3 \leftrightarrow 4 ) + ( 1 \leftrightarrow 2, 3 \leftrightarrow 4  )   \,. 
\end{align} 
As a cross-check of our SMEFT computation we have also calculated the squared matrix elements for the tree-level processes $h \to b \bar b q \bar q$, $h \to b \bar b b \bar b$ and $h \to b \bar b gg$ within the SM. Our~findings agree with~(A.8), (A.9), (A.11), (A.13), (A.14) and (A.16) of~\cite{DelDuca:2015zqa}. 

In the case of the $h (p_h) \to b (p_1) \bar b (p_2)$ transition, the first non-zero contribution to the interference  with an insertion of $Q_{bG}$ and the corresponding SM contribution arises at~${\cal O} (y_b^2  \hspace{0.25mm} \alpha_s^3  \hspace{0.25mm} C_{bG})$. We find
\beq \label{eq:hbb2loop}
{\cal S}_{b \bar b}^{(2)}= 2  \hspace{0.25mm} {\rm Re}  \, \Big \langle {\cal M}_{y_b, b \bar b}^{(0)} \, \Big | \, {\cal M}_{C_{bG}, b \bar b}^{(2)} \Big  \rangle = -\frac{\alpha_s}{ 256 \hspace{0.25mm} \pi^3}  \, N_\epsilon^2 \hspace{0.5mm} {\cal S}_{b \bar bg}^{ (0)} \hspace{0.5mm} m_h^2    \hspace{0.5mm}  \gamma_{bG} \hspace{0.5mm} \big ( 7 + 2 \hspace{0.125mm} L \big ) \,, 
\eeq
where~${\cal M}_{C_{bG}, b \bar b}^{(2)}$ denotes the relevant two-loop $h \to b \bar b$ matrix element in the SMEFT and the expressions for~${\cal S}_{b \bar bg}^{ (0)}$,  $N_\epsilon$ and $\gamma_{bG}$ have been given in~(\ref{eq:MEybdgLO}), (\ref{eq:Neps}) and (\ref{eq:ADMs}), respectively. 

\section{Numerical impact of $\bm{C_{HG}}$ and $\bm{C_{3G}}$ }
\label{app:others}

In this appendix we assess the numerical impact of the dimension-six operators $Q_{HG}$ and~$Q_{3G}$ on the $h \to b \bar b$ decay and the $pp \to Zh$ production process.  Since we show below that these operators contribute to less than a percent at the level of total rates once existing experimental limits on the relevant Wilson coefficients are taken into account, the effects of $Q_{HG}$ and $Q_{3G}$ have been neglected in the phenomenological analysis carried out in the main part  of this work.  

\begin{figure}[t!]
\begin{center}
\includegraphics[height=0.175\textwidth]{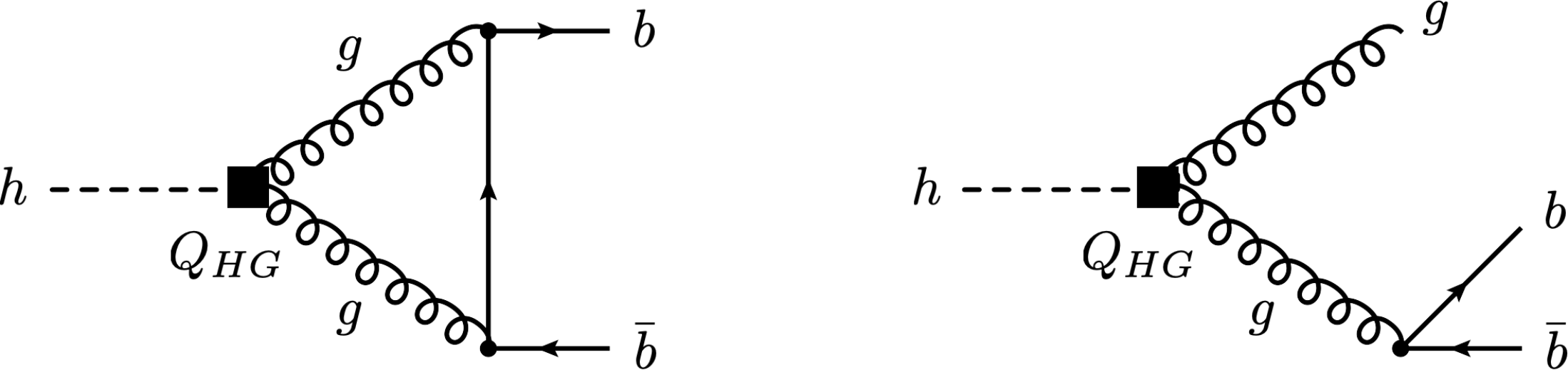}
\end{center}
\vspace{0mm} 
\caption{\label{fig:diagramsQHGdecay} Examples of SMEFT contributions to the inclusive decay rate of $h \to b \bar b$ with an insertion of~$Q_{HG}$~(black square). The  left (right) diagram represents a one-loop (tree-level) contribution to the $h \to b \bar b$ ($h \to b \bar b g$) transition.  }
\end{figure}

The leading correction to the inclusive $h \to b \bar b$ decay rate proportional to the Wilson coefficient~$C_{HG}$ has been calculated for massive bottom quarks in~\cite{Gauld:2016kuu}.  Example diagrams are shown in Figure~\ref{fig:diagramsQHGdecay}. Employing the operator basis~(\ref{eq:operators}) and working to leading power in $m_b$ one finds at NNLO in QCD the simple expression 
\beq \label{eq:GammaSMEFTNNLOHG}
\Gamma ( h \to b \bar b )_{\rm SMEFT}^{{\rm NNLO}, HG} =  \left ( \frac{\alpha_s}{\pi} \right )^2 \left [ \frac{19}{3} - 2 \hspace{0.125mm} \zeta_2 +   \frac{1}{3} \ln^2 \left ( \frac{\bar m_b^2}{m_h^2} \right ) \right ] c_{HG} \hspace{0.5mm}  \Gamma ( h \to b \bar b )_{\rm SM}^{\rm LO}  \,,
\eeq
where the definition of the Wilson coefficient $c_{HG}$ can be found in~(\ref{eq:cHGdef}). A recent global  fit~\cite{Ellis:2020unq} to the SMEFT including~34 dimension-six operators  reports the following marginalised 95\%~CL bound  
\beq \label{eq:cHG95CL}
c_{HG} \in [-0.09, 0.06] \,, 
\eeq
on the relevant Wilson~coefficient (cf.~Table~6~of~the latter work). Using this limit together with the input parameters given at the beginning of Section~\ref{sec:numerics} it  follows from~(\ref{eq:GammaSMEFTNNLOHG}) and~(\ref{eq:cHG95CL})  that at~95\%~CL the~$Q_{HG}$ contribution to the inclusive $h \to b \bar b$ decay rate lies within
\beq \label{eq:GammaHG95CL}
\frac{\Gamma ( h \to b \bar b )_{\rm SMEFT}^{{\rm NNLO}, HG}}{\Gamma ( h \to b \bar b )_{\rm SM}^{\rm LO}  } \in [-2.7, 1.7] \cdot 10^{-3} \,. 
\eeq
This result indicates that SMEFT effects  arising from $Q_{HG}$ are phenomenologically  irrelevant in the case of the fully differential $h \to b \bar b$ decay rate. This justifies that we have neglected such corrections in the main part of this article. 

Operator insertions of $Q_{3G}$ induce tree-level contributions to $h \to b \bar b gg$ and one-loop corrections to  $h \to b \bar b g$. After interfering these channels with the corresponding SM amplitudes and integrating over the  four- and three-particle phase space, respectively, the combination of these two types of contributions leads to a N$^3$LO correction  to the inclusive $h \to b \bar b$ decay rate. We  write  the sum of these real and virtual corrections as
\beq \label{eq:GammaSMEFTNNNLO3G}
\Gamma ( h \to b \bar b )_{\rm SMEFT}^{{{\rm N}^3{\rm LO}}, 3G} =  N_{3G}^{\rm dec} \left ( \frac{\alpha_s}{\pi} \right )^3 \frac{m_h^2}{v^2} \hspace{0.5mm} c_{3G} \hspace{0.5mm}  \Gamma ( h \to b \bar b )_{\rm SM}^{\rm LO}  \,,
\eeq
where we have defined 
\beq \label{eq:c3Gdef}
c_{3G} = \frac{v^2}{\Lambda^2} \hspace{0.5mm} C_{3G} \,.
\eeq
The  marginalised 95\%~CL~limit on the relevant Wilson~coefficient reads~\cite{Ellis:2020unq}\footnote{The large negative values of $c_{3G}$ found in the work~\cite{Ellis:2020unq} can be traced back to the discrepancy between the measured $t \bar t$ differential cross section~\cite{CMS:2018htd} and the state-of-the-art SM prediction~\cite{Czakon:2017wor} at large values of the top-antitop invariant mass~$m_{t \bar t}$. Analyses of multijet data~\cite{Krauss:2016ely,Hirschi:2018etq,Goldouzian:2020wdq} suggest bounds of $|c_{3G}| \lesssim 0.2$.  }
\beq \label{eq:c3G95CL}
c_{3G} \in [-12.5, -4.1] \,.
\eeq
Plugging~(\ref{eq:c3G95CL}) into~(\ref{eq:GammaSMEFTNNNLO3G}) then leads to  
\beq \label{eq:Gamma3G95CL}
\frac{\Gamma ( h \to b \bar b )_{\rm SMEFT}^{{{\rm N}^3{\rm LO}}, 3G}}{\Gamma ( h \to b \bar b )_{\rm SM}^{\rm LO}  } \in [-0.15, -0.05] \cdot 10^{-3} \hspace{0.25mm} N_{3G}^{\rm dec}  \,. 
\eeq
By performing an explicit calculation of the tree-level contributions to $h \to b \bar b gg$ and the one-loop corrections to  $h \to b \bar b g$ associated to insertions of $Q_{3G}$, we find for the unknown constant $N_{3G}^{\rm dec}$ introduced  in~(\ref{eq:GammaSMEFTNNNLO3G}) the numerial value $N_{3G}^{\rm dec}  = 2.23$. This implies that the relative corrections associated to the operator~$Q_{3G}$ do not even reach the level of a  permille. Neglecting these corrections as done in our SMEFT analysis of the fully differential $h \to b \bar b$ decay rate is therefore  fully~justified from a phenomenological point of view. 

\begin{figure}[t!]
\begin{center}
\includegraphics[height=0.19\textwidth]{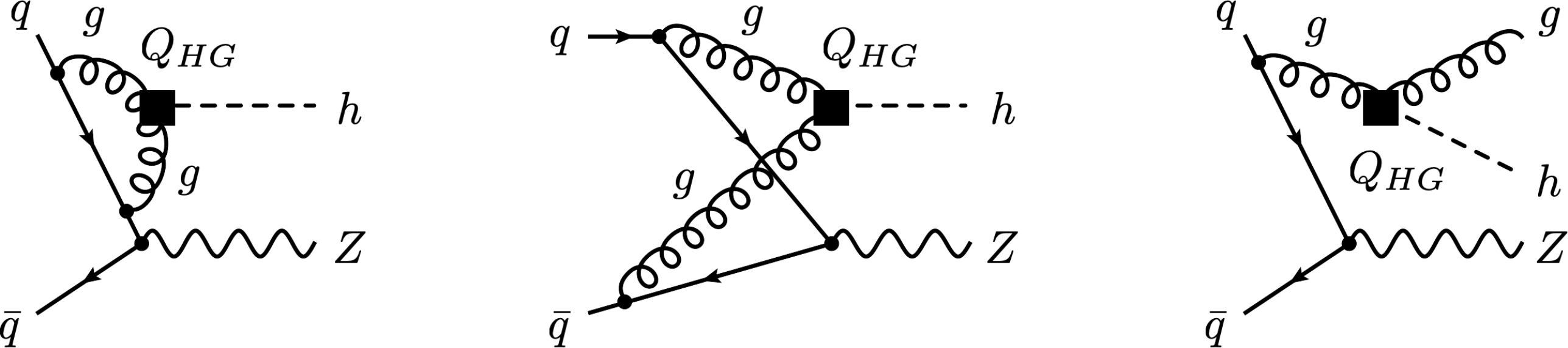}
\end{center}
\vspace{-1mm} 
\caption{\label{fig:diagramsQHGproduction}  Examples of SMEFT contributions to  $Zh$ production involving an insertion of~$Q_{HG}$~(black square). The  left and middle (right) diagram represent(s) a one-loop (tree-level) contribution to the $q \bar q \to Zh$ ($q \bar q \to Zhg$) process.  }
\end{figure}

The leading corrections to $Zh$ production associated to the operator $Q_{HG}$ result from  the Feynman graphs displayed in Figure~\ref{fig:diagramsQHGproduction}.  These types of diagrams have been calculated in the context of the SM in~\cite{Brein:2011vx} working in the limit of infinite top-quark mass. Using the results of the latter work one can write the~${\cal O} (\alpha_s^2 \hspace{0.5mm} C_{HG})$ corrections to the inclusive~$Zh$~production cross section in the following way 
\beq \label{eq:prodQHG}
\sigma ( p p \to Zh )_{\rm SMEFT}^{{\rm NNLO}, HG} = 3 \left ( \frac{\alpha_s}{\pi} \right )^2 \hspace{0.5mm}   \delta \hspace{0.5mm} c_{HG} \, \sigma ( p p \to Zh )_{\rm SM}^{\rm LO} \,, 
\eeq
where $ \sigma ( p p \to Zh )_{\rm \bibliography{hbb}
SM}^{\rm LO}$ is the LO  cross section and  $\delta$ encodes the sum of the QCD corrections denoted by $V_{\rm I}$ and $R_{\rm I}$ in~\cite{Brein:2011vx}  with a factor of $(\alpha_s/\pi)^2$ stripped off. We add that we have calculated the relevant radiative corrections that give rise to $V_{\rm I}$ and $R_{\rm I}$ finding agreement with the latter publication (see also~\cite{Brein:2012ne,Harlander:2018yio}). Notice that an  expression analogous to~(\ref{eq:prodQHG}) also holds for the differential cross section. From Figure~6 of the paper~\cite{Brein:2011vx} one finds that at the LHC one has~$\delta = 10.7$ for the measured mass of the Higgs boson. This corresponds to a $1.4\%$ correction in the SM.   Using the limit~(\ref{eq:cHG95CL}) in~(\ref{eq:prodQHG}) we obtain 
\beq \label{eq:ppZhHG}
\frac{\sigma ( p p \to Zh )_{\rm SMEFT}^{{{\rm NNLO}}, HG}}{ \sigma ( p p \to Zh )_{\rm SM}^{\rm LO} } \in [-3.9, 2.4] \cdot 10^{-3} \,,
\eeq
at  95\%~CL. This numerical result shows that it is an excellent approximation to neglect contributions due to $Q_{HG}$ in the calculation of $Zh$ production observables. 

Insertions of the operator $Q_{3G}$ induce tree-level contributions to $q \bar q \to Z h gg$ and one-loop corrections to  $q \bar q \to Z h  g$. Interfering the SMEFT with the corresponding SM   amplitudes and integrating over the relevant phase spaces, one obtains a  N$^3$LO correction  to the inclusive~$Zh$~production cross section. Using naive dimensional analysis, we write this correction as follows 
\beq \label{eq:sigmaSMEFTNNNLO3G}
\sigma ( p p \to Zh )_{\rm SMEFT}^{{{\rm N}^3{\rm LO}}, 3G} =  N_{3G}^{\rm prod} \left ( \frac{\alpha_s}{\pi} \right )^3  \hspace{0.5mm} c_{3G} \, \sigma ( p p \to Zh )_{\rm SM}^{\rm LO}   \,,
\eeq
where $N_{3G}^{\rm prod}$ is an undetermined numerical factor  which is naively of~${\cal O}(1)$. From~(\ref{eq:c3G95CL}) and~(\ref{eq:sigmaSMEFTNNNLO3G}) it follows that 
\beq \label{eq:sigma3G95CL}
\frac{\sigma ( p p \to Zh )_{\rm SMEFT}^{{{\rm N}^3{\rm LO}}, 3G} }{ \sigma ( p p \to Zh )_{\rm SM}^{\rm LO}    } \in [-0.58, -0.19] \cdot 10^{-3} \hspace{0.25mm} N_{3G}^{\rm prod}  \,.
\eeq
For $N_{3G}^{\rm prod}= {\cal O} (10)$ this 95\%~CL bound implies a relative correction of a few permille to the inclusive  $pp\to Zh$ cross section. In practice, the ${{\rm N}^3{\rm LO}}$ corrections~(\ref{eq:sigmaSMEFTNNNLO3G}) can therefore be neglected. 

\end{appendix}

%\bibliographystyle{JHEPmod}
%\bibliography{hbb}

%\end{document}

\providecommand{\href}[2]{#2}\begingroup\raggedright\endgroup

\end{document}